\documentclass[journal]{IEEEtran}[12pt]
\ifCLASSINFOpdf
\usepackage[pdftex]{graphicx}
\graphicspath{{../pdf/}{../jpeg/}}
\DeclareGraphicsExtensions{.pdf,.jpeg,.png}
\else
\usepackage[dvips]{graphicx}
\graphicspath{{../eps/}}
\DeclareGraphicsExtensions{.eps}
\fi\usepackage{graphicx}

\usepackage{graphics}
\usepackage{epsfig}
\usepackage{epstopdf}
\usepackage{stfloats}
\usepackage[cmex10]{amsmath}
\usepackage{algorithmic}
\usepackage[ruled]{algorithm2e}
\SetKwRepeat{Do}{do}{while}
\usepackage{tablefootnote}
\usepackage{array}
\usepackage{mdwmath}
\usepackage{mdwtab}
\usepackage{graphicx}
\usepackage{subfigure}
\usepackage{color}
\usepackage{amsfonts,amssymb}
\usepackage{hyperref} 
\usepackage{multirow}
\usepackage{diagbox}
\usepackage{caption}
\usepackage{bbding}
\usepackage{amsfonts,amsthm,array} 
\usepackage{makecell}
\usepackage{lineno} 

\usepackage{float}  

\begin{document}
	
	
	\title{Aerial Relay to Achieve Covertness and Secrecy	\thanks{Manuscript received.}}
	
	\author{Jiacheng~Jiang,
	Hongjiang~Lei, ~\IEEEmembership{Senior Member,~IEEE,}
	Ki-Hong~Park, ~\IEEEmembership{Senior Member,~IEEE,}\\
	Gaofeng~Pan, ~\IEEEmembership{Senior Member,~IEEE,}
	and~Mohamed-Slim~Alouini, ~\IEEEmembership{Fellow,~IEEE}
	\thanks{This work was supported by the National Natural Science Foundation of China under Grant 62171031 and 61971080. (Corresponding author: \textit{Hongjiang Lei}.)}
	\thanks{Jiacheng~Jiang and Hongjiang~Lei are with the School of Communications and Information Engineering, Chongqing University of Posts and Telecommunications, Chongqing 400065, China, also with Chongqing Key Lab of Mobile Communications Technology, Chongqing 400065, China (e-mail: cquptjjc@163.com, leihj@cqupt.edu.cn).}
	\thanks{Ki-Hong~Park and Mohamed-Slim~Alouini are with the CEMSE Division, King Abdullah University of Science and Technology (KAUST), Thuwal 23955-6900, Saudi Arabia (e-mail: kihong.park@kaust.edu.sa, slim.alouini@kaust.edu.sa).}
	\thanks{Gaofeng~Pan is with the School of Cyberspace Science and Technology, Beijing Institute of Technology, Beijing 100081, China (e-mail: gfpan@bit.edu.cn).}
	}

\maketitle

\begin{abstract}
In this work, we investigate a delay-tolerant covert and secure communication framework relayed by an uncrewed aerial vehicle (UAV).
In this framework, a legitimate UAV serves as an aerial relay to facilitate communication when the direct link between the terrestrial transmitter and receiver is blocked.
Additionally, the UAV also acts as a friendly jammer and adopts a random jamming transmit power technique to assist the covert communications.
In order to achieve optimal trade-offs between these dual functionalities, a phase transition mechanism has been developed, regulated through a phase-switching factor.
Given the uncertainty of the malicious nodes' positions, we formulate a robust fractional programming optimization problem aimed at maximizing the covert and secure energy efficiency by jointly optimizing the UAV's trajectory, the transmitter's power, and the phase-switching factor.
An alternating optimization-based algorithm is then proposed to solve the fractional programming problem.
To ensure low computational complexity, we solve the phase-switching factor sub-problem with a primal-dual search-based algorithm and the others with successive convex approximation-based algorithms.
The effectiveness of the proposed algorithm is validated through numerical results.
\end{abstract}



\section{Introduction}
\label{sec:Introduction}

\subsection{Background and Related Works}
\label{sec:BackgroundandRelatedWorks}

On account of the advantages of maneuverability and high mobility, uncrewed aerial vehicles (UAVs) were used in wireless communication applications, including intelligent logistics, precision agriculture, data relaying, as well as disaster rescue \cite{ZengY2016Mag, KimH2018Mag, LinX2018Mag}.
Specifically, the UAV-assisted network is a promising way to establish an emergency communication framework for crucial rescue when natural disasters occur \cite{ZhaoN2019WC}.
In contrast to traditional terrestrial wireless communication systems, UAV-assisted wireless communication systems can own a high probability that the air-to-ground (A2G) channels are line-of-sight (LoS) through position or trajectory design \cite{ZengY2016Mag, WuQ2019WC}.
Nevertheless, the high probability of LoS links also provides convenience for malicious nodes, which exposes confidential information to risky situations.
Consequently, how to transmit classified information in UAV-assisted communication systems has become a crucial issue in future investigations \cite{DuoB2021ChinaCom, ChenX2022Netw}.
As a matter of course, physical layer security (PLS) and covert communications (CC), as prospective technologies for ensuring communication secrecy and covertness respectively, offer a viable solution to this challenge \cite{ChenX2020Netw, JiangX2021WC, YangB2022Netw}.

\emph{1) PLS-assisted UAV Communication Systems:}
One of the challenges of information security in UAV communication systems is that the openness of wireless medium makes systems more vulnerable to eavesdropping \cite{HamamrehJM2019Survey}.
Existing investigations indicate that the UAV communication systems can be protected from eavesdropping by cooperative jamming \cite{CaiY2018JSAC, WangW2021JSAC, ZhangR2021TWC, WangZ2022JCN} and position or trajectory design \cite{CuiM2018TVT, ZhangG2019TWC, DuoB2020TVT}.
In particular, for cooperative jamming, a dual-UAV-enabled secure communication system wherein a UAV serves as a base station (BS) with another cooperative UAV that acts as a friendly jammer to realize the secrecy rate maximization was investigated in \cite{CaiY2018JSAC} and \cite{WangW2021JSAC}.
In addition, a dual-UAV-assisted secure data collection system that considered more practical propulsion energy consumption constraint was studied, and the total average secrecy rate (ASR) was maximized by optimizing the scheduling, transmit power, trajectory and velocity in \cite{ZhangR2021TWC}.
In \cite{WangZ2022JCN}, authors investigated a dual-UAV-assisted cognitive relay system comprising a secondary UAV relay and jammer, maximizing the ASR through joint optimization of robust trajectories and transmit power for both UAVs.
Besides, in \cite{CuiM2018TVT}, the authors considered a UAV secure communication system with multiple terrestrial eavesdropping nodes, and the ASR was sub-optimally maximized by designing the transmit power and trajectory of UAV.
Furthermore, in \cite{ZhangG2019TWC}, a full-duplex (FD) UAV secure communication system was considered to investigate how to maximize the ASR by jointly optimizing the UAV's trajectory and the transmit power of the transmitter.
Extensions based on \cite{CuiM2018TVT} and \cite{ZhangG2019TWC}, in \cite{DuoB2020TVT}, the secure energy efficiency (SEE) of the UAV was maximized by jointly designing the trajectory and transmit/jamming powers of UAV with FD scheme in the UAV secure communication system.
Additionally, studies on secure communication systems in UAV relay networks were conducted in \cite{SunG2022TCOM} and \cite{KongL2024IoT}.
Specifically, \cite{SunG2022TCOM} proposed a novel aerial relay system based on collaborative beamforming via a UAV-enabled virtual antenna array, achieving secure and energy-efficient communication for remote ground users, and \cite{KongL2024IoT} introduced two cognitive UAV relay selection schemes, namely, a non-interference aware relay selection scheme and an interference aware relay selection scheme.

\emph{2) CC-assisted UAV Communication Systems:} 
Unlike PLS, CC focuses on guaranteeing that the communication process is undetectable by malicious nodes \cite{BashBA2015Mag, ChenX2023Survey, YangF2024CJA}.
In order to achieve communication covertness while getting significant performance, how to suppress malicious nodes effectively becomes a primary objective.
Through existing studies, the purpose can be achieved via random transmit power \cite{YanS2019TIFS}, random noise \cite{LiM2023TVT, ZhouX2019TSP}, and friendly jamming \cite{ChenX2021TVT, LeiH2024TAES}.
Specifically, for random transmit power, in \cite{YanS2019TIFS}, the authors investigated the covert throughput (CT) of a delay-intolerant system with finite block length under situations of fixed and random transmit power, respectively, and the results demonstrate that the CT has significantly improved when transmit power is stochastic.
For random noise, \cite{LiM2023TVT} studied a UAV relaying system that adopted the amplify-and-forward (AF) model with stochastic noise assist.
Moreover, a UAV communication network with random noise help was considered in \cite{ZhouX2019TSP}, and the average covert rate (ACR) was maximized via jointly optimizing the trajectory and transmit power of the UAV.
Besides, \cite{ChenX2021TVT} and \cite{LeiH2024TAES} aim to research the influence of friendly jamming on covert performance.
Significantly, the conclusions of \cite{LeiH2024TAES} show that the UAV acting as the jamming device can efficiently improve the minimum ACR of the multi-user aerial communication system, and the detection error probability (DEP) during the scenario with monitoring nodes equipped with multiple antennas was formulated.
In addition to the above, a recent study \cite{WangC2023TCOM} investigated a novel communication system using UAV-mounted intelligent reflecting surfaces (IRS),
\cite{WangH2024TVT} studied a covert UAV relaying system with a fixed-trajectory aerial warden,
and \cite{YangF2024TVT} examined countermeasures against multiple active monitoring nodes.

\begin{table*}
	\centering
	\caption{{Differences Between Our Work and Existing Works}}
	\label{Table 1}
	\resizebox{0.8\textwidth}{!}
	{
	\begin{tabular}{ c | c | c | c | c | c | c | c }
		\Xhline{1.2pt}
		\multirow{2}{*}{\textbf{Reference}}&	\multirow{2}{*}{\textbf{Technology}}&	\multirow{2}{*}{\textbf{Role of UAV(s)}}&	\multirow{2}{*}{\makecell[c]{\textbf{Location of Malicious} \\ \textbf{Node(s)}}}&	\multicolumn{3}{c|}{\textbf{Optimization Variables}}&	\multirow{2}{*}{\textbf{Objective}}\\
		\cline{5-7}
		&	&	&	&	\textbf{trajectory}&	\makecell[c]{\textbf{transmit} \\ \textbf{power}}&	\makecell[c]{\textbf{switching} \\ \textbf{factor}}&	\\
		\hline
		\cite{ZhangG2022TGCN}&	-&	relay&	-&	\checkmark&	\checkmark&	-&	throughput\\
		\hline
		\cite{CaiY2018JSAC}&	\multirow{9}{*}{PLS}&	\multirow{2}{*}{BS and friendly jammer}&	\multirow{8}{*}{imperfect}&	\checkmark&	-&	-&	\multirow{6}{*}{ASR}\\
		\cline{1-1} \cline{5-7}
		\cite{WangW2021JSAC}&	&	&	&	\checkmark&	-&	\checkmark&	\\
		\cline{1-1} \cline{3-3} \cline{5-7}
		\cite{ZhangR2021TWC}&	&	receiver and jammer&	&	\checkmark&	\checkmark&	-&	\\
		\cline{1-1} \cline{3-3} \cline{5-7}
		\cite{WangZ2022JCN}&	&	cognitive relay and jammer&	&	\checkmark&	\checkmark&	-&	\\
		\cline{1-1} \cline{3-3} \cline{5-7}
		\cite{CuiM2018TVT}&	&	\multirow{2}{*}{BS}&	&	\checkmark&	\checkmark&	-&	\\
		\cline{1-1} \cline{5-7}
		\cite{ZhangG2019TWC}&	&	&	&	\checkmark&	\checkmark&	-&	\\
		\cline{1-1} \cline{3-3} \cline{5-8}
		\cite{DuoB2020TVT}&	&	BS and jammer&	&	\checkmark&	\checkmark&	-&	\multirow{2}{*}{SEE}\\
		\cline{1-1} \cline{3-3} \cline{5-7}
		\cite{SunG2022TCOM}&	&	\multirow{2}{*}{relay}&	&	\checkmark&	-&	-&	\\
		\cline{1-1} \cline{4-7} \cline{8-8}
		\cite{KongL2024IoT}&	&	&	perfect&	-&	-&	-&	-\\
		\cline{1-3} \cline{4-7}  \cline{8-8}
		\cite{YanS2019TIFS}&	\multirow{8}{*}{CC}&	-&	\multirow{3}{*}{imperfect}&	-&	\checkmark&	-&	CT\\
		\cline{1-1} \cline{3-3} \cline{5-8}
		\cite{LiM2023TVT}&	&	relay&	&	\checkmark&	\checkmark&	-&	\multirow{7}{*}{ACR}\\
		\cline{1-1} \cline{3-3} \cline{5-7}
		\cite{ZhouX2019TSP}&	&	\multirow{3}{*}{BS}&	&	\checkmark&	\checkmark&	-&	\\
		\cline{1-1} \cline{4-4} \cline{5-7}
		\cite{ChenX2021TVT}&	&	&	perfect&	-&	\checkmark&	-&	\\
		\cline{1-1} \cline{4-4} \cline{5-7}
		\cite{JiangX2021TVT}&	&	&	\multirow{4}{*}{imperfect}&	\checkmark&	\checkmark&	-&	\\
		\cline{1-1} \cline{3-3} \cline{5-7}
		\cite{LeiH2024TAES}&	&	BS and friendly jammer&	&	\checkmark&	\checkmark&	-&	\\
		\cline{1-1} \cline{3-3} \cline{5-7}
		\cite{WangC2023TCOM}&	&	UAV-IRS&	&	-&	\checkmark&	-&	\\
		\cline{1-1} \cline{3-3} \cline{5-7}
		\cite{WangH2024TVT}&	&	relay&	&	\checkmark&	\checkmark&	-&	\\
		\cline{1-4} \cline{5-8}
		\cite{WangHM2020TCOM}&	\multirow{5}{*}{PLS\&CC}&	-&	perfect&	-&	\checkmark&	-&	\multirow{2}{*}{C\&ST}\\
		\cline{1-1} \cline{3-4} \cline{5-7}
		\cite{WangC2022TWC}&	&	-&	\multirow{2}{*}{imperfect}&	-&	\checkmark&	-&	\\
		\cline{1-1} \cline{3-3} \cline{5-8}
		\cite{LiuP2023TVT}&	&	BS&	&	\checkmark&	\checkmark&	-&	\multirow{2}{*}{AC\&SR}\\
		\cline{1-1} \cline{3-4} \cline{5-7}
		\cite{WangX2024TWC}&	&	-&	perfect&	-&	\checkmark&	-&	\\
		\cline{1-1} \cline{3-4} \cline{5-7} \cline{8-8}
		our work&	&	relay and jammer&	imperfect&	\checkmark&	\checkmark&	\checkmark&	C\&SEE\\
		\Xhline{1.2pt}
	\end{tabular}
}
\end{table*}

\subsection{Motivations and Contributions}
\label{sec:Motivations and Contributions}

Based on the above, most existing studies on UAV communication systems have focused on either PLS or CC in isolation, achieving secrecy (via PLS) or covertness (via CC) individually, but failing to integrate both capabilities.
However, the distinction between these two technologies is too critical to be negligible: \textit{CC ensures that the transmission remains undetected by wardens, while PLS ensures that the content remains confidential from eavesdroppers.}
Against this backdrop, applying either PLS or CC technology in isolation can create vulnerabilities that allow other types of threats to emerge.
For instance, when a base station employs conventional artificial noise (AN) to counter eavesdroppers during data transmission, the excessive transmit power may lead to unintended exposure of the communication process to the warden.
To address these complexities, researchers have proposed a joint PLS and CC scheme.
Specifically, \cite{WangHM2020TCOM} proposed a multi-hop scheme by optimizing the coding rates, transmit power, and required number of hops to counter UAV surveillance and achieve covertness and secrecy, while \cite{WangC2022TWC} introduced a broadcast communication paradigm by designing the beamforming matrix and the covariance matrix of AN to achieve both covertness and secrecy.
Additionally, the authors in \cite{LiuP2023TVT} developed a pioneering joint model that integrates information-theoretic covertness and secrecy in a UAV-assisted finite blocklength transmission system.
Moreover, \cite{WangX2024TWC} studied a body-centric Internet of Things (IoT) communication system with information secrecy features based on PLS and CC techniques, derived closed-form expressions for the secrecy outage probability and secrecy rate in PLS, and the detection error probability and covert rate in CC by adopting the Alternate Rician Shadowed fading model, and proposed a selection algorithm for PLS and CC techniques to reach a high transmission performance state while ensuring communication covertness and secrecy.
Nevertheless, existing studies in \cite{WangHM2020TCOM}, \cite{WangC2022TWC}, \cite{LiuP2023TVT} and \cite{WangX2024TWC}, which investigate covert and secure communication in the terrestrial multi-hop network, the ground base station scenario, the IoT network and the aerial base station scenario, respectively, are not universally applicable to aerial relay systems.
Moreover, these works primarily focus on maximizing the covert and secure throughput (C\&ST) or the average covert and secure rate (AC\&SR), while overlooking the energy-constrained nature of UAV communication systems, which is a critical practical limitation.
Building on this foundation, our work explores a delay-tolerant UAV-relayed covert and secure communication framework.
Table \ref{Table 1} illustrates the differences and hence the advantages of our work, comparing with state-of-the-art work.
Specifically, the main contributions are summarized as follows:

\begin{itemize}
	
	\item We consider a legitimate UAV serving as an aerial relay to facilitate communication when the direct link between the transmitter and receiver is obstructed.
	Meanwhile, a terrestrial warden near the transmitter attempts to monitor the transmission, and a terrestrial eavesdropper near the receiver seeks to intercept the classified data.
	To ensure covertness during the communication process, the UAV also acts as a cooperative jammer and employs a random jamming transmit power technique to disrupt the warden's monitoring efforts.
	To balance the switching between these two roles, a phase-switching scheme has been proposed and is controlled by a phase-switching factor.
	Subsequently, given the uncertainty in the positions of the malicious nodes, a robust fractional programming (FP) optimization problem is formulated to maximize the lower bound of covert and secure energy efficiency (C\&SEE) by jointly optimizing the UAV's trajectory, the transmitter's power, and the phase-switching factor.
	
	\item Since the non-convexity of the proposed problem, an efficient alternating optimization (AO)-based algorithm is proposed, which decomposes the original problem into three sub-problems to address this intractable non-convex FP problem.
	To develop a low-complexity optimization algorithm, a primal-dual search-based algorithm (PDSA) is introduced to efficiently solve the sub-problem related to the phase-switching factor, while the remaining sub-problems are addressed using the successive convex approximation (SCA) method.
	
	\item In contrast to the perfect position information assumed in \cite{ChenX2021TVT}, this work considers practical and realistic scenarios where the positions of malicious nodes are assumed to be uncertain.
	This uncertainty increases the complexity of the optimization problem, making it more challenging to solve.
	
\end{itemize}

\subsection{Organization}
\label{sec:Organization}

The remainder of this paper is organized as follows.
Section \ref{sec:SystemModelandProblemFormulation} presents the system model and formulates the corresponding optimization problem.
An AO-based algorithm is established in Section \ref{sec:ProblemSolution} to solve the problem formulated in Section \ref{sec:SystemModelandProblemFormulation}.
In Section \ref{sec:Simulation}, the superiority and effectiveness of our proposed scheme are demonstrated through numerical results.
Finally, Section \ref{sec:Conclusion} concludes this work.

\section{System Model and Problem Formulation}
\label{sec:SystemModelandProblemFormulation}

\begin{figure}[t]
	\centering		
	\includegraphics[width = 0.4  \textwidth]{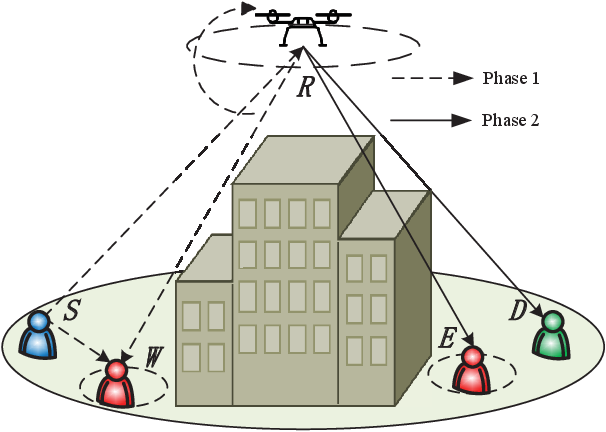}
    \caption{Delay-tolerant UAV relayed covert and secure communication model.}
    \label{fig_model}
\end{figure}

\subsection{System Model}
\label{sec:SystemModel}
As shown in Fig. \ref{fig_model}, consider a delay-tolerant aerial relaying system with high demand for covertness and secrecy, where the direct link between the source (${S}$) and the destination (${D}$) is blocked.
${S}$ transmits the confidential data to ${D}$ with the help of an aerial relay (${R}$).
At the same time, a terrestrial warden (${W}$) near ${S}$ tries to decide whether ${S}$ transmits or not, and a terrestrial eavesdropper (${E}$) near ${D}$ wants to eavesdrop the confidential data sent by ${S}$ to ${D}$ through ${R}$'s relaying.
In this system, all the terrestrial nodes are equipped with a single antenna.
Assume that ${R}$ serves as a relay node within a period of $T$.
The entire process is divided into two phases with durations of ${\alpha T}$ and ${\left( {1 - \alpha } \right)T}$, respectively, where $\alpha \in \left( {0,1} \right)$ denotes the phase-switching factor between two phases.
In the first phase, ${R}$, which is equipped with a receive antenna and a transmit antenna and employs the decode-and-forward (DF) model, adopts the FD mode for receiving and jamming to ensure the covertness of the system.
Subsequently, ${R}$ retransmits the confidential data to ${D}$ securely in the second phase.
It is worth noting that in the first phase, since ${R}$ is operating in the FD manner, self-interference cancellation (SIC) has to be applied.
Nevertheless, since the SIC at ${R}$ cannot be canceled entirely in practice, the residual self-interference channel (RSIC) is considered in this work.
Similar to \cite{DuoB2020TVT}, the RSIC ${{h_{RR}}}$ is modeled as a Rayleigh fading channel, which means ${h_{RR}} \sim \mathcal{CN}\left( {0,\psi } \right)$, where $\psi$ denotes the SIC level.

For the concise and clear trajectory described, the duration of the first phase ${\alpha T}$ is equally divided into ${N_1}$ time slots as ${\delta _{t,1}} = \frac{{\alpha T}}{{{N_1}}}$, and the duration of the second phase ${\left( {1 - \alpha } \right)T}$ is equally divided into ${N_2}$ time slots as ${\delta _{t,2}} = \frac{{\left( {1 - \alpha } \right)T}}{{{N_2}}}$. When ${\delta _{t,k}},k \in \left\{ {1,2} \right\}$ is small enough, the position of ${R}$ is approximately fixed for each time slot \cite{CuiM2018TVT}. Without loss of generality, all nodes are described in the Cartesian coordinate system, and the horizontal location of ${R}$ is denoted as ${{\bf{q}}_R}\left( n \right) = {\left[ {{x_R}\left( n \right),{y_R}\left( n \right)} \right]^T},n \in {\mathcal{N}}$, where ${\cal N} = \left\{ {1, \cdots ,N} \right\}$ and $N = {N_1} + {N_2}$. For all the ground nodes, their coordinate positions are expressed as ${{\bf{q}}_S} = {\left[ {{x_S},{y_S}} \right]^T}$, ${{\bf{q}}_D} = {\left[ {{x_D},{y_D}} \right]^T}$, ${{\bf{q}}_W} = {\left[ {{x_W},{y_W}} \right]^T}$, and ${{\bf{q}}_E} = {\left[ {{x_E},{y_E}} \right]^T}$, respectively. To guarantee the robustness of the communication system, the practical situation where the malicious nodes' position is uncertain is assumed \cite{YangF2024TVT}, \cite{WangC2023TCOM}. According to this assumption, the positions of ${W}$ and ${E}$ are denoted as ${{\bf{q}}_W} = {{\bf{\hat q}}_W} + \Delta {{\bf{q}}_W}$ and ${{\bf{q}}_E} = {{\bf{\hat q}}_E} + \Delta {{\bf{q}}_E}$, respectively, where ${{\bf{\hat q}}_j}$, $j \in \left\{ {W,E} \right\}$ are the estimated positions made by ${R}$-mounted cameras or radars \cite{ZhouX2019TSP}, and $\Delta {{\bf{q}}_j}$ denotes the estimation error, with ${\left| {\Delta {{\bf{q}}_j}} \right|^2} \le r_j^2$, where $r_j$ is the degree of estimation error for the corresponding malicious node.

In this work, ${R}$ operates at a fixed altitude $H$, and according to \cite{LinX2018Mag}, the appropriate setting of the UAV flight height can make the LoS probability approach 1. Thus, it is assumed that the A2G links are LoS \cite{ZhangR2021TWC, DuoB2020TVT}. Subsequently, the channel coefficients for ${R}$ in two phases are expressed as
\begin{subequations}
	\begin{align}
		&{h_{SR}}\left( n \right) = \sqrt {\frac{{{\beta _0}}}{{{{\left\| {{{\bf{q}}_R}\left( n \right) - {{\bf{q}}_S}} \right\|}^2} + {H^2}}}}, \\
		&{h_{RW}}\left( n \right) = \sqrt {\frac{{{\beta _0}}}{{{{\left\| {{{\bf{q}}_R}\left( n \right) - {{\bf{q}}_W}} \right\|}^2} + {H^2}}}},
		\label{phase1abR}
	\end{align}
\end{subequations}
and
\begin{subequations}
	\begin{align}
		&{h_{RD}}\left( n \right) = \sqrt {\frac{{{\beta _0}}}{{{{\left\| {{{\bf{q}}_R}\left( n \right) - {{\bf{q}}_D}} \right\|}^2} + {H^2}}}}, \\
		&{h_{RE}}\left( n \right) = \sqrt {\frac{{{\beta _0}}}{{{{\left\| {{{\bf{q}}_R}\left( n \right) - {{\bf{q}}_E}} \right\|}^2} + {H^2}}}},
		\label{phase2abR}
	\end{align}
\end{subequations}
respectively, where ${{\beta _0}}$ denotes the channel power gain at the reference distance. For the ground-to-ground (G2G) link, the quasi-static Rayleigh model is utilized, with the channel coefficient from $S$ to $W$ expressed as
\begin{equation}
	{h_{SW}} = \sqrt {\frac{{{\beta _0}}}{{{{\left\| {{{\bf{q}}_S} - {{\bf{q}}_W}} \right\|}^\eta }}}} \zeta,
	\label{phase1StoW}
\end{equation}
where $\zeta$ follows a complex Gaussian distribution with zero mean and unit variance, and $\eta > 2$ represents the path-loss exponent.

\subsection{Covert Constraint}
\label{sec:CovertConstraint}

\begin{figure*}[t]
	\begin{equation}
		y_W^i\left( n \right) = \left\{ {\begin{array}{*{20}{c}}
				{\sqrt {P_R^J} {h_{RW}}\left( n \right)x_J^i\left( n \right) + n_W^i\left( n \right)},&{{{\cal H}_0}}\\
				{\sqrt {{P_S}\left( n \right)} {h_{SW}}x_S^i\left( n \right) + \sqrt {P_R^J} {h_{RW}}\left( n \right)x_J^i\left( n \right) + n_W^i\left( n \right)},&{{{\cal H}_1}}
		\end{array}} \right.
		\label{binary hypothesis}
	\end{equation}
	\hrulefill
\end{figure*}

Similar to \cite{YanS2019TIFS}-\cite{YangF2024TVT}, a binary hypothesis testing problem is modeled by $W$ to determine whether $S$ is transmitting or not.
At the same time, it is assumed that the exact locations of $S$ and $R$ are known to $W$, which is the worst-case scenario in CC.
Then, for $W$, the binary hypothesis testing in the $n$-th time slot is modeled as (\ref{binary hypothesis}), shown at the top of this page, where ${{\mathcal{H}_0}}$ and ${{\mathcal{H}_1}}$ represent the hypotheses that $S$ is or is not transmitting, ${{P_S}{\left( n \right)}}$ and ${P_R^J}$ represent the transmit power of $S$ and the jamming power of $R$ during the first phase, $x_J^i{\left( n \right)}$ and $x_S^i{\left( n \right)}$ are the transmitted signals that follow a complex Gaussian distribution with zero mean and unit variance, and $n_W^i\left( n \right)$ is the complex additive Gaussian noise with power ${\sigma ^2}$ at $W$.
In addition, a uniform jammer model is adopted in this work \cite{MaoH2024IoT, ZhouX2021JSAC, SobersTV2017TWC}, which means that ${P_R^J}$ is a random variable following the uniform distribution over the interval $\left[ {0,\hat P_R^J} \right]$.
Hence, the probability density function of ${P_R^J}$ is given by
\begin{equation}
	{f_{P_R^J}}\left( x \right) = \left\{ {\begin{array}{*{20}{l}}
			{\frac{1}{{\hat P_R^J}},}&{0 \le x \le \hat P_R^J,}\\
			{0,}&{{\rm{otherwise.}}}
	\end{array}} \right.
	\label{PDF}
\end{equation}
Under the worst-case scenario in CC, the distribution information of ${P_R^J}$ in (\ref{PDF}) is also assumed to be perfectly known to $W$.

Similar to \cite{LiM2023TVT, MaoH2024IoT}, it is assumed that $W$ observes an infinite number of samples to make a decision.
Therefore, the optimal decision rule at $W$ is expressed as
\begin{equation}
	{T_W}\left( n \right) = \mathop {\lim }\limits_{I \to \infty } \frac{1}{I}\sum\limits_{i = 1}^I {{{\left| {y_W^i\left( n \right)} \right|}^2}} \mathop {\mathop  \gtrless \limits_{{\mathcal{D}_0}} }\limits^{{\mathcal{D}_1}} {\tau \left( n \right)},
	\label{rule}
\end{equation}
where ${\mathcal{D}_0}$ and ${\mathcal{D}_1}$ represent $W$'s decision biased towards ${{\cal H}_0}$ and ${{\cal H}_1}$, respectively, ${\tau \left( n \right)}$ represents the detection threshold, which can be optimized to minimize the DEP of $W$, and ${T_W}\left( n \right)$ is the average received power, which can be simplified as
\begin{equation}
	{T_W}\left( n \right) = \left\{ {\begin{array}{*{20}{l}}
			{P_R^J{{\left| {{h_{RW}}\left( n \right)} \right|}^2} + {\sigma ^2},}&{{{\cal H}_0},}\\
			{{P_S}\left( n \right){{\left| {{\bar h_{SW}}} \right|}^2} + P_R^J{{\left| {{h_{RW}}\left( n \right)} \right|}^2} + {\sigma ^2},}&{{{\cal H}_1},}
	\end{array}} \right.
	\label{T_W}
\end{equation}
where ${\left| {{{\bar h}_{SW}}} \right|^2} = {\mathbb{E}_\zeta}\left( {{{\left| {{h_{SW}}} \right|}^2}} \right) = \frac{{{\beta _0}}}{{{{\left\| {{{\bf{q}}_S} - {{\bf{q}}_W}} \right\|}^\eta }}}$.

The detection performance of $W$ is typically modeled as DEP in this work \cite{YanS2019TIFS}-\cite{YangF2024TVT}.
It is composed of false alarm probability (FAP) and miss detection probability (MDP), which are expressed as ${{\mathop{\rm P}\nolimits} _{{\rm{FA}}}}\left( n \right) = \Pr \left( {{{\cal D}_1}|{{\cal H}_0}} \right)$ and ${{\mathop{\rm P}\nolimits} _{{\rm{MD}}}}\left( n \right) = \Pr \left( {{{\cal D}_0}|{{\cal H}_1}} \right)$, respectively \cite{ChenX2023Survey}.
According to (\ref{rule}) and (\ref{T_W}), the FAP at $W$ is calculated as
\begin{equation}
	\begin{aligned}
		{{\rm{P}}_{{\rm{FA}}}}\left( n \right) &= \Pr \left\{ {P_R^J \ge \frac{{\tau \left( n \right) - {\sigma ^2}}}{{{{\left| {{h_{RW}}\left( n \right)} \right|}^2}}}} \right\} \\
		&= \int_{\max \left( {0,\frac{{\tau \left( n \right) - {\sigma ^2}}}{{{{\left| {{h_{RW}}\left( n \right)} \right|}^2}}}} \right)}^{\hat P_R^J} {{f_{P_R^J}}\left( x \right)dx} \\
		&= \left\{ {\begin{array}{*{20}{l}}
				{1,}&{\tau \left( n \right) < {\sigma ^2},}\\
				{1 - \frac{{\tau \left( n \right) - {\sigma ^2}}}{{\hat P_R^J{{\left| {{h_{RW}}\left( n \right)} \right|}^2}}},}&{{\sigma ^2} \le \tau \left( n \right) < {z_1},}\\
				{0,}&{{z_1} \le \tau \left( n \right),}
		\end{array}} \right.
		\label{}
	\end{aligned}
\end{equation}
where ${z_1} = \hat P_R^J{\left| {{h_{RW}}\left( n \right)} \right|^2} + {\sigma ^2}$.
Similarly, the MDP at $W$ is given by
\begin{equation}
	\begin{aligned}
		{{\rm{P}}_{{\rm{MD}}}}\left( n \right) &= \Pr \left\{ {P_R^J \le \frac{{\tau \left( n \right) - {P_S}\left( n \right){{\left| {{\bar h_{SW}}} \right|}^2} - {\sigma ^2}}}{{{{\left| {{h_{RW}}\left( n \right)} \right|}^2}}}} \right\} \\
		&= \int_0^{\min \left( {\hat P_R^J,\frac{{\tau \left( n \right) - {P_S}\left( n \right){{\left| {{\bar h_{SW}}} \right|}^2} - {\sigma ^2}}}{{{{\left| {{h_{RW}}\left( n \right)} \right|}^2}}}} \right)} {{f_{P_R^J}}\left( x \right)dx} \\
		&= \left\{ {\begin{array}{*{20}{l}}
				{0,}&{\tau \left( n \right) < {z_2},}\\
				{\frac{{\tau \left( n \right) - {P_S}\left( n \right){{\left| {{\bar h_{SW}}} \right|}^2} - {\sigma ^2}}}{{\hat P_R^J{{\left| {{h_{RW}}\left( n \right)} \right|}^2}}},}&{{z_2} \le \tau \left( n \right) < {z_3},}\\
				{1,}&{{z_3} \le \tau \left( n \right),}
		\end{array}} \right.
		\label{}
	\end{aligned}
\end{equation}
where ${z_2} = {P_S}\left( n \right){\left| {{\bar h_{SW}}} \right|^2} + {\sigma ^2}$ and ${z_3} = {P_S}\left( n \right){\left| {{\bar h_{SW}}} \right|^2} + \hat P_R^J{\left| {{h_{RW}}\left( n \right)} \right|^2} + {\sigma ^2}$.
It is worth noting that if ${P_S}\left( n \right){\left| {{\bar h_{SW}}} \right|^2} \ge \hat P_R^J{\left| {{h_{RW}}\left( n \right)} \right|^2}$, zero DEP is achieved at $W$ by setting $\tau \left( n \right) = {z_2}$.
Therefore, the constraint ${P_S}\left( n \right){\left| {{\bar h_{SW}}} \right|^2} < \hat P_R^J{\left| {{h_{RW}}\left( n \right)} \right|^2}$ should be satisfied.
Subsequently, the DEP of $W$ is expressed as
\begin{equation}
	\begin{aligned}
		{\xi _W}\left( n \right) &= {{\rm{P}}_{{\rm{FA}}}}\left( n \right) + {{\rm{P}}_{{\rm{MD}}}}\left( n \right) \\
		&= \left\{ {\begin{array}{*{20}{l}}
				{1,}&{\tau \left( n \right) < {\sigma ^2},}\\
				{1 - \frac{{\tau \left( n \right) - {\sigma ^2}}}{{\hat P_R^J{{\left| {{h_{RW}}\left( n \right)} \right|}^2}}},}&{{\sigma ^2} \le \tau \left( n \right) < {z_2},}\\
				{1 - \frac{{{P_S}\left( n \right){{\left| {{\bar h_{SW}}} \right|}^2}}}{{\hat P_R^J{{\left| {{h_{RW}}\left( n \right)} \right|}^2}}},}&{{z_2} \le \tau \left( n \right) < {z_1},}\\
				{\frac{{\tau \left( n \right) - {P_S}\left( n \right){{\left| {{\bar h_{SW}}} \right|}^2} - {\sigma ^2}}}{{\hat P_R^J{{\left| {{h_{RW}}\left( n \right)} \right|}^2}}},}&{{z_1} \le \tau \left( n \right) < {z_3},}\\
				{1,}&{{z_3} \le \tau \left( n \right).}
		\end{array}} \right.
		\label{}
	\end{aligned}
\end{equation}
Since the goal of $W$ is to minimize the DEP, a proper detection threshold should be designed.
It is clear that ${\xi _W}\left( n \right)$ monotonically decreases with $\tau \left( n \right)$ for ${{\sigma ^2} \le \tau \left( n \right) < {z_2}}$, monotonically increases with $\tau \left( n \right)$ for ${{z_1} \le \tau \left( n \right) < {z_3}}$, and is a constant over interval ${{z_2} \le \tau \left( n \right) < {z_1}}$.
Hence, considering that ${\xi _W}\left( n \right)$ is a continuous function of $\tau \left( n \right)$, the minimum DEP ${\xi^* _W} \left( n \right)$ at $W$ is achieved by setting the optimal detection threshold $\tau^* \left( n \right)$ within the interval $\left[ {{z_2},{z_1}} \right)$, which is given by
\begin{equation}
	\xi _W^*\left( n \right) = 1 - \frac{{{P_S}\left( n \right){{\left| {{\bar h_{SW}}} \right|}^2}}}{{\hat P_R^J{{\left| {{h_{RW}}\left( n \right)} \right|}^2}}}.
	\label{}
\end{equation}

To ensure covertness during the communication process, ${\xi^* _W}\left( n \right) \ge 1 - \varepsilon$ should be satisfied in the first phase, where $0 < \varepsilon < 1$ is a sufficiently small positive value that represents the quality of covert service.
Finally, the covert constraint is expressed as
\begin{equation}
	{P_S}\left( n \right){\left| {{\bar h_{SW}}} \right|^2} \le \varepsilon \hat P_R^J{\left| {{h_{RW}}\left( n \right)} \right|^2}.
	\label{}
\end{equation}

\subsection{Problem Formulation}

According to the description in Section \ref{sec:SystemModel} and the DF protocol, the achievable covert and secure rates in the two phases are modeled as
\begin{equation}
	\begin{aligned}
		{R_1}\left( n \right) &= {{\mathbb{E}}_{h_{RR}}}\left( {B{{\log }_2}\left( {1 + \frac{{{P_S}\left( n \right){{\left| {{h_{SR}}\left( n \right)} \right|}^2}}}{{P_R^J{{\left| {{h_{RR}}} \right|}^2} + {\sigma ^2}}}} \right)} \right) \\
		&\mathop  \ge \limits^{\left( a \right)} B{\log _2}\left( {1 + \frac{{{P_S}\left( n \right){{\left| {{h_{SR}}\left( n \right)} \right|}^2}}}{{P_R^J\psi  + {\sigma ^2}}}} \right)  \triangleq  {R_1^{\sec }\left( n \right)},
		\label{phase1rate}
	\end{aligned}
\end{equation}
and
\begin{equation}
	{R_2}\left( n \right) = {\left[ {{R_D}\left( n \right) - {R_E}\left( n \right)} \right]^ + },
	\label{phase2rate}
\end{equation}
respectively, where ${\left[ \cdot \right]^ + } = \max \left( {0, \cdot } \right)$, step $\left( a \right)$ is taken using Jensen's inequality \cite{BoydS2004Book}, $B$ presents the channel bandwidth in hertz (Hz), ${R_D}\left( n \right) = B{\log _2}\left( {1 + \frac{{{P_R}\left( n \right){{\left| {{h_{RD}}\left( n \right)} \right|}^2}}}{{{\sigma ^2}}}} \right)$ and ${R_E}\left( n \right) = B{\log _2}\left( {1 + \frac{{{P_R}\left( n \right){{\left| {{h_{RE}}\left( n \right)} \right|}^2}}}{{{\sigma ^2}}}} \right)$. Then, the AC\&SR is stated as
\begin{equation}
	{R_{ave}} = \min \left( {{\phi _1}\sum\limits_{n = 1}^{{N_1}} {R_1^{\sec }\left( n \right)} ,{\phi _2}\sum\limits_{n = {N_1} + 1}^N {{R_2}\left( n \right)} } \right),
	\label{Averate}
\end{equation}
where ${\phi _1} = \frac{\alpha }{{{N_1}}}$, ${\phi _2} = \frac{{\left( {1 - \alpha } \right)}}{{{N_2}}}$. Nevertheless, due to the presence of the estimation error in (\ref{Averate}), attention must be paid to the stochasticity of ${R_{ave}}$. To deal with this issue, the triangle inequality is employed to obtain a lower bound of ${R_{ave}}$, which is shown as
\begin{equation}
	\bar R_{ave}^{\sec } = \min \left( {{\phi _1}\sum\limits_{n = 1}^{{N_1}} {R_1^{\sec }\left( n \right)} ,{\phi _2}\sum\limits_{n = {N_1} + 1}^N {{{\left[ {R_2^{\sec }\left( n \right)} \right]}^ + }} } \right),
	\label{Averatelowerbound}
\end{equation}
here
\begin{equation}
	R_2^{\sec }\left( n \right) = {{R_D}\left( n \right) - B{{\log }_2}\left( {1 + \frac{{{P_R}\left( n \right){{\left| {{{\hat h}_{RE}}\left( n \right)} \right|}^2}}}{{{\sigma ^2}}}} \right)},
	\label{phase2ratelowerbound}
\end{equation}
and ${\hat h_{RE}}\left( n \right) = \sqrt {\frac{{{\beta _0}}}{{{{\left( {\left\| {{{\bf{q}}_R}\left( n \right) - {{{\bf{\hat q}}}_E}} \right\| - {r_E}} \right)}^2} + {H^2}}}}$ based on the triangle inequality, where ${\hat h_{RE}}\left( n \right) \ge {h_{RE}}\left( n \right)$ holds.

\begin{figure*}[t]
	\begin{equation}
		{P_C}\left( n \right) = {P_0}\left( {1 + \frac{{3{{\left\| {{{\bf{v}}_R}\left( n \right)} \right\|}^2}}}{{U_{tip}^2}}} \right) + {P_1}\sqrt {\left( {\sqrt {1 + \frac{{{{\left\| {{{\bf{v}}_R}\left( n \right)} \right\|}^4}}}{{4v_0^4}}}  - \frac{{{{\left\| {{{\bf{v}}_R}\left( n \right)} \right\|}^2}}}{{2v_0^2}}} \right)}  + \frac{1}{2}{d_0}\rho cS{\left\| {{{\bf{v}}_R}\left( n \right)} \right\|^3}
		\label{propulsionpower}
	\end{equation}
	\hrulefill
\end{figure*}

As derived in \cite{ZengY2019TWC}, for a rotary-wing UAV, the propulsion power is expressed as (\ref{propulsionpower}), shown at the top of the next page, where ${P_0}$ and ${P_1}$ represent the blade profile and induced power in hovering status, respectively, ${U_{tip}}$ signifies the tip speed of the rotor blade, ${v_0}$ is the mean rotor induced velocity in hover, ${d_0}$, $\rho$, $c$ and $S$ are known as the fuselage drag ratio, air density, rotor solidity and rotor disc area, respectively, and ${{{\bf{v}}_R}\left( n \right)}$ expressed as
\begin{equation}
	{{\bf{v}}_R}\left( n \right) = \frac{1}{{{\delta _{t,k}}}}\left( {{{\bf{q}}_R}\left( n \right) - {{\bf{q}}_R}\left( {n - 1} \right)} \right),n \in {{\mathcal{N}}_k},
\end{equation}
where ${{\cal N}_1} = \left\{ {1, \cdots ,{N_1}} \right\}$ and ${{\cal N}_2} = \left\{ {{N_1} + 1, \cdots ,N} \right\}$.
When $\left| x \right| \ll 1$, via first-order Taylor approximation, $\sqrt {1 + x}  \approx 1 + \frac{1}{2}x$ is satisfied.
Similarly, when $\left\| {{{\bf{v}}_R}\left( n \right)} \right\| \gg {v_0}$, (\ref{propulsionpower}) is approximately transformed as
\begin{equation}
	\begin{aligned}
		P_C^{\sec }\left( n \right) =& {P_0}\left( {1 + \frac{{3{{\left\| {{{\bf{v}}_R}\left( n \right)} \right\|}^2}}}{{U_{tip}^2}}} \right) \\
		&+ \frac{{{P_1}{v_0}}}{{\left\| {{{\bf{v}}_R}\left( n \right)} \right\|}} + \frac{1}{2}{d_0}\rho cS{\left\| {{{\bf{v}}_R}\left( n \right)} \right\|^3},
		\label{approximatepower}
	\end{aligned}
\end{equation}
where the typical plot of the relationship between (\ref{propulsionpower}) and (\ref{approximatepower}) is given by \cite{ZengY2019TWC}.
According to the Fig. 2 of \cite{ZengY2019TWC}, $P_C^{\sec }\left( n \right) \ge {P_C}\left( n \right)$ is satisfied.
Additionally, the power utilized for transmission and jamming is much smaller than that for flight, so that can be neglected.
Subsequently, an upper bound of the total energy consumption of $R$ in communication process is expressed as
\begin{equation}
	{E_{sum}} = {\delta _{t,1}}\sum\limits_{n = 1}^{{N_1}} {P_C^{\sec }\left( n \right)}  + {\delta _{t,2}}\sum\limits_{n = {N_1} + 1}^N {P_C^{\sec }\left( n \right)}.
	\label{energyconsumption}
\end{equation}

In this work, C\&SEE is defined as the ratio of AC\&SR and total energy consumption of $R$.
Based on this definition, a tractable lower bound of C\&SEE of $R$ is formulated as
\begin{equation}
	\varphi  = \frac{{\bar R_{ave}^{\sec }}}{{{E_{sum}}}}.
\end{equation}
To improve the energy efficiency of the system while ensuring the covertness and secrecy, the maximization of $\varphi$ is pursued.
Notably, the parameter $\alpha$ functions as a dual effect\footnote{The phase-switching factor $\alpha$ plays a crucial role in balancing the relaying and cooperative jamming functionalities in $R$. When $\alpha$ is either too small or too large, it shortens the duration of either phase 1 or phase 2, respectively. A small $\alpha$ restricts the time available for $R$ to effectively use trajectory planning for jamming, while a large $\alpha$ limits the duration of secure relaying in phase 2. Consequently, inappropriate values of $\alpha$ can undermine the system's covert or secure performance.},
directly influencing the trade-off between relaying and cooperative jamming functionalities in $R$, which in turn affects the system's C\&SEE.
Subsequently, a robust optimization problem concerning the phase-switching factor $\alpha$, the transmit power in each phase ${{\bf{P}}_S} = \left\{ {{P_S}\left( n \right),n \in {{\mathcal{N}}_1}} \right\}$, ${{\bf{P}}_{R,2}} = \left\{ {{P_R}\left( n \right),n \in {{\mathcal{N}}_2}} \right\}$, and the trajectory of $R$, denoted as ${{\bf{Q}}_R} = \left\{ {{{\bf{q}}_R}\left( n \right),\forall n} \right\}$, is formulated as
\begin{subequations}
	\begin{align}
		\mathcal{P}_{1}:\;&\mathop {\max }\limits_{{{\bf{P}}_S},{{\bf{P}}_{R,2}},{{\bf{Q}}_R},\alpha} \varphi \label{P1a}\\
		{\mathrm{s.t.}}\;
		&{P_S}\left( n \right){\left| {{\bar h_{SW}}} \right|^2} \le \varepsilon \hat P_R^J{\left| {{h_{RW}}\left( n \right)} \right|^2}, n \in {{\mathcal{N}}_1}, \label{P1b}\\
		&0 \le {P_S}\left( n \right) \le {P_{S,\max }},n \in {{\mathcal{N}}_1}, \label{P1c}\\
		&0 \le {P_R}\left( n \right) \le {P_{R,\max }},n \in {{\mathcal{N}}_2}, \label{P1d}\\
		&{{\bf{q}}_R}\left( 1 \right) = {{\bf{q}}^I},{{\bf{q}}_R}\left( N \right) = {{\bf{q}}^F}, \label{P1e}\\
		&\left\| {{{\bf{q}}_R}\left( n \right) - {{\bf{q}}_R}\left( {n - 1} \right)} \right\| \le {\delta _{t,1}}{V_{\max }},n \in {{\mathcal{N}}_1}, \label{P1f}\\
		&\left\| {{{\bf{q}}_R}\left( n \right) - {{\bf{q}}_R}\left( {n - 1} \right)} \right\| \le {\delta _{t,2}}{V_{\max }},n \in {{\mathcal{N}}_2}, \label{P1g}\\
		&0 < \alpha  < 1, \label{P1h}
	\end{align}
\end{subequations}
where
(\ref{P1b}) is the covert constraint,
(\ref{P1c}) and (\ref{P1d}) are the peak power constraint of $S$ in phase one and $R$ in phase two, respectively, where ${P_{S,\max }}$ and ${P_{R,\max }}$ signify the maximum transmit power,
(\ref{P1e}) denotes the constraint on the take-off and landing positions of $R$,
${{\bf{q}}^I}$ and ${{\bf{q}}^F}$ are the take-off and landing positions of $R$, respectively,
(\ref{P1f}) and (\ref{P1g}) depict the maximum flight distance between adjacent time slots in phase one and phase two, respectively,
and
${V_{\max }}$ denotes the maximum velocity of $R$.

\section{Problem Solution}
\label{sec:ProblemSolution}

Problem $\mathcal{P}_{1}$ is a multivariate coupled non-convex FP problem and includes the position estimation error of $W$.
As derived in \cite{CuiM2018TVT}, removing the operator ${\left[  \cdot  \right]^ + }$, the optimal value is also the same.
Subsequently, using the Dinkelbach method \cite{CrouzeixJP} and the triangle inequality, the original FP problem is transformed into the following problem to obtain the approximate solution, namely
\begin{subequations}
	\begin{align}
		\mathcal{P}_{2}:\;&\mathop {\max }\limits_{{{\bf{P}}_S},{{\bf{P}}_{R,2}},{{\bf{Q}}_R},\alpha } R_{ave}^{\sec } - {\varphi ^{\left( l \right)}}{E_{sum}} \label{P2a}\\
		{\mathrm{s.t.}}\;
		&{P_S}\left( n \right){\left| {\bar h_{SW}^{\sec }} \right|^2} \le \varepsilon \hat P_R^J{\left| {h_{RW}^{\sec }\left( n \right)} \right|^2},n \in {{\cal N}_1}, \label{P2b}\\
		&(\textrm{\ref{P1c}}) - (\textrm{\ref{P1h}}), \label{P2c}
	\end{align}
\end{subequations}
where $\bar h_{SW}^{\sec } = \sqrt {\frac{{{\beta _0}}}{{{{\left| {\left\| {{{\bf{q}}_S} - {{{\bf{\hat q}}}_W}} \right\| - {r_W}} \right|}^\eta }}}}  \ge {\bar h_{SW}}$,
$h_{RW}^{\sec }\left( n \right) = \sqrt {\frac{{{\beta _0}}}{{{{\left( {\left\| {{{\bf{q}}_R}\left( n \right) - {{{\bf{\hat q}}}_W}} \right\| + {r_W}} \right)}^2} + {H^2}}}}  \le {h_{RW}}\left( n \right)$,
${\varphi ^{\left( l \right)}} = \frac{{\bar R_{ave}^{\sec ,\left( l \right)}}}{{E_{sum}^{\left( l \right)}}}$ is the value of C\&SEE obtained in $l$-th iteration
and
$R_{ave}^{\sec } = \min \left( {{\phi _1}\sum\limits_{n = 1}^{{N_1}} {R_1^{\sec }\left( n \right)} ,{\phi _2}\sum\limits_{n = {N_1} + 1}^N {R_2^{\sec }\left( n \right)} } \right)$.
Two factors in $\mathcal{P}_{2}$ must be noticed.
The first is that the covert constraint (\ref{P2b}) is a non-convex constraint; the other is that the objective function, which is highly complicated with respect to the variables ${{\bf{P}}_S}$, ${{\bf{P}}_{R,2}}$, ${{\bf{Q}}_R}$, and $\alpha$ is non-concave.
For the two challenges mentioned above, an AO-based algorithm is presented in the following subsection.

\subsection{AO-Based Algorithm}

In this section, we present an efficient AO-based algorithm that strives to obtain a high-quality sub-optimal solution to problem $\mathcal{P}_{1}$.
Our approach involves solving three sub-problems alternately in order to optimize the phase-switching factor $\alpha$, the transmit power ${{\bf{P}}_S}$ and ${{\bf{P}}_{R,2}}$, as well as the $R$'s trajectory ${{\bf{Q}}_R}$, thereby addressing problem $\mathcal{P}_{1}$ sub-optimally.

\begin{algorithm}[t]
	\caption{Primal-Dual Search Based Algorithm}
	\label{A1}
	Calculate ${\beta _1}$, ${\beta _2}$, ${\rho _1}$, ${\rho _2}$, ${\hat \alpha _1}$, ${\hat \alpha _2}$ and ${\hat \alpha _3}$, respectively;
	
	\eIf{${\hat \alpha _1} \in \left( {{\beta _1},1 - {\beta _2}} \right)$}
	{
		\eIf{${\hat \alpha _1}{\rho _1} \le \left( {1 - {{\hat \alpha }_1}} \right){\rho _2}$}
		{${\alpha ^*} = {\hat \alpha _1}$.}
		{	
			\eIf{${\hat \alpha _2} \in \left( {{\beta _1},{{\hat \alpha }_1}} \right)$}
			{
				\eIf{${\hat \alpha _2} \ge {\hat \alpha _3}$}
				{${\alpha ^*} = {\hat \alpha _2}$.}
				{${\alpha ^*} = {\hat \alpha _3}$.}
			}
			{
				\eIf{${\beta _1} \ge {\hat \alpha _3}$}
				{${\alpha ^*} = {\beta _1}$.}
				{${\alpha ^*} = {\hat \alpha _3}$.}
			}
		}
	}
	{
		\eIf{${\hat \alpha _1} \le {\beta _1}$}
		{${\alpha ^*} = {\beta _1}$.}
		{
			\eIf{$\left( {1 - {\beta _2}} \right){\rho _1} \le {\beta _2}{\rho _2}$}
			{${\alpha ^*} = 1 - {\beta _2}$.}
			{	
				\eIf{${\hat \alpha _2} \in \left[ {{\beta _1},1 - {\beta _2}} \right)$}
				{
					\eIf{${\hat \alpha _2} \ge {\hat \alpha _3}$}
					{${\alpha ^*} = {\hat \alpha _2}$.}
					{${\alpha ^*} = {\hat \alpha _3}$.}
				}
				{
					\eIf{${\beta _1} \ge {\hat \alpha _3}$}
					{${\alpha ^*} = {\beta _1}$.}
					{
						\eIf{${\hat \alpha _3} \le {1 - {\beta _2}}$}
						{${\alpha ^*} = {\hat \alpha _3}$.}
						{${\alpha ^*} = {1 - {\beta _2}}$.}
					}
				}
			}
		}
	}
	
	\KwOut{${\alpha ^*}$.}
\end{algorithm}

\emph{1) Phase-switching factor:} For fixed transmit power ${{\bf{P}}_S}$ and ${{\bf{P}}_{R,2}}$, and $R$'s trajectory ${{\bf{Q}}_R}$, problem $\mathcal{P}_{2}$ is reduced to
\begin{subequations}
	\begin{align}
		\mathcal{P}_{2.1.1}:\;&\mathop {\max }\limits_{\alpha } R_{ave}^{\sec } - {\varphi ^{\left( l \right)}}{E_{sum}} \label{P3.1.1a}\\
		{\mathrm{s.t.}}\;
		&(\textrm{\ref{P1f}}) - (\textrm{\ref{P1h}}), \label{P3.1.1b}
	\end{align}
\end{subequations}
which is a non-convex optimization problem.
By introducing the slack variable ${\omega _1}$, an equivalent form of $\mathcal{P}_{2.1.1}$ is obtained as
\begin{subequations}
	\begin{align}
		\mathcal{P}_{2.1.2}:\;&\mathop {\max }\limits_{\alpha ,{\omega _1}} \; {\omega _1} - {\varphi ^{\left( l \right)}}{E_{sum}} \label{P3.1.2a}\\
		{\mathrm{s.t.}}\;
		& - {\omega _1} + \alpha {\rho _1} \ge 0, \label{P3.1.2b}\\
		& - {\omega _1} + \left( {1 - \alpha } \right){\rho _2} \ge 0, \label{P3.1.2c}\\
		&\;\alpha - {\beta _1} \ge 0, \label{P3.1.2d}\\
		& - \alpha - {\beta _2} + 1 \ge 0, \label{P3.1.2e}
	\end{align}
\end{subequations}
which is a standard convex optimization problem that satisfies Slater's condition, where ${\rho _1} = \frac{1}{{{N_1}}}\sum\limits_{n = 1}^{{N_1}} {R_1^{\sec }\left( n \right)}$,
${\rho _2} = \frac{1}{{{N_2}}}\sum\limits_{n = {N_1} + 1}^N {R_2^{\sec }\left( n \right)}$, 
and
${\beta _k} = \frac{{{N_k}\mathop {\max }\limits_{n \in {{\mathcal{N}}_k}} \left\| {{{\bf{q}}_R}\left( n \right) - {{\bf{q}}_R}\left( {n - 1} \right)} \right\|}}{{T{V_{\max }}}}$.

The KKT conditions of $\mathcal{P}_{2.1.2}$ is formed as
\begin{subequations}
	\begin{align}
		& - {\varphi ^{\left( l \right)}}\frac{{d{E_{sum}}\left( {{\alpha ^*}} \right)}}{{d\alpha }} + \lambda _1^*{\rho _1} - \lambda _2^*{\rho _2} + \lambda _3^* - \lambda _4^* = 0, \label{stationarity1}\\
		&1 - \lambda _1^* - \lambda _2^* = 0,\lambda _1^* \ge 0,\lambda _2^* \ge 0,\lambda _3^* \ge 0,\lambda _4^* \ge 0, \label{stationarity2}\\
		&\lambda _1^*\left( { - \omega _1^* + {\alpha ^*}{\rho _1}} \right) = 0,\lambda _2^*\left( { - \omega _1^* + \left( {1 - {\alpha ^*}} \right){\rho _2}} \right) = 0, \label{complementarity1}\\
		&\lambda _3^*\left( {{\alpha ^*} - {\beta _1}} \right) = 0,\lambda _4^*\left( { - {\alpha ^*} - {\beta _2} + 1} \right) = 0, \label{complementarity2}\\
		& - \omega _1^* + {\alpha ^*}{\rho _1} \ge 0, - \omega _1^* + \left( {1 - {\alpha ^*}} \right){\rho _2} \ge 0, \label{primaryfeasible1}\\
		&{\alpha ^*} - {\beta _1} \ge 0, - {\alpha ^*} - {\beta _2} + 1 \ge 0,
		\label{primaryfeasible2}
	\end{align}
\end{subequations}
where $\lambda _1^*$, $\lambda _2^*$, $\lambda _3^*$ and $\lambda _4^*$ are the Lagrange multipliers.
It is challenging to derive an analytical expression for ${\alpha ^*}$ directly from the above conditions; therefore, a novel algorithm based on the primal-dual search method, PDSA, has been proposed, and the derivation details are shown as follows.

\setcounter{equation}{29}
\begin{figure*}[t]
	\begin{equation}
		B{\log _2}\left( {{\sigma ^2} + {P_R}\left( n \right){{\left| {{{\hat h}_{RE}}\left( n \right)} \right|}^2}} \right) \le \underbrace {B{{\log }_2}\left( {{\sigma ^2} + P_R^{\left( l \right)}\left( n \right){{\left| {{{\hat h}_{RE}}\left( n \right)} \right|}^2}} \right) + \frac{{B{{\left| {{{\hat h}_{RE}}\left( n \right)} \right|}^2}\left( {{P_R}\left( n \right) - P_R^{\left( l \right)}\left( n \right)} \right)}}{{\left( {{\sigma ^2} + P_R^{\left( l \right)}\left( n \right){{\left| {{{\hat h}_{RE}}\left( n \right)} \right|}^2}} \right)\ln \left( 2 \right)}}}_{ \buildrel \Delta \over = K\left( n \right)}
		\label{TPSCA}
	\end{equation}
	\hrulefill
\end{figure*}
\setcounter{equation}{27}

By substituting (\ref{stationarity2}) into (\ref{stationarity1}), the stationarity condition is equally transformed into
\begin{equation}
	- {\varphi ^{\left( l \right)}}\frac{{d{E_{sum}}\left( {{\alpha ^*}} \right)}}{{d\alpha }} + {\rho _1} - \lambda _2^*\left( {{\rho _1} + {\rho _2}} \right) + \lambda _3^* - \lambda _4^* = 0.
	\label{stationarity}
\end{equation}
To search for the optimal value ${\alpha ^*}$, the following factors should be noted.

\begin{enumerate}
	
	\item[a)] Since ${{E_{sum}}\left( {{\alpha }} \right)}$ is convex function with respect to ${\alpha }$, it is not difficult to know that $\frac{{d{E_{sum}}\left( {{\alpha }} \right)}}{{d\alpha }}$ is monotonically increasing with respect to ${\alpha }$.
	
	\item[b)] The value of ${\hat \alpha _1}$ which meets the equation $ - {\varphi ^{\left( l \right)}}\frac{{d{E_{sum}}\left( {{{\hat \alpha }_1}} \right)}}{{d\alpha }} + {\rho _1} = 0$ can be obtained by the bisection search method.
	
	\item[c)] The value of ${\hat \alpha _2}$ is obtained by the equation ${\hat \alpha _2}{\rho _1} = \left( {1 - {{\hat \alpha }_2}} \right){\rho _2}$.
	
	\item[d)] The value of ${\hat \alpha _3}$ which meets the equation $ - {\varphi ^{\left( l \right)}}\frac{{d{E_{sum}}\left( {\hat \alpha _3} \right)}}{{d\alpha }} - {\rho _2} = 0$ is obtained by the bisection search method.
	
	\item[e)] Based on a), b) and d), ${\hat \alpha _1} > {\hat \alpha _3}$ is satisfied.
	
	\item[f)] According to the objective function of $\mathcal{P}_{2.1.2}$, it is known that if the feasibility conditions (\ref{primaryfeasible1}) are met, a larger value for $\omega _1^*$ is preferable.
	
\end{enumerate}

\noindent
When ${\hat \alpha _1} \in \left( {{\beta _1},1 - {\beta _2}} \right)$, depending on different cases, one can obtain

\begin{enumerate}
	
	\item[i)] If ${\hat \alpha _1}{\rho _1} \le \left( {1 - {{\hat \alpha }_1}} \right){\rho _2}$ is satisfied, ${\hat \alpha _1}$ is the optimal value of $\alpha$.
	This is because when ${\alpha ^*} = {\hat \alpha _1}$, the corresponding solutions $\omega _1^*$, $\lambda _1^*$, $\lambda _2^*$, $\lambda _3^*$ and $\lambda _4^*$ that satisfy the conditions can be found.
	
	\item[ii)] When ${\hat \alpha _1}{\rho _1} > \left( {1 - {{\hat \alpha }_1}} \right){\rho _2}$,
	if ${\hat \alpha _2} \in \left( {{\beta _1},{{\hat \alpha }_1}} \right)$ and ${\hat \alpha _2} \ge {\hat \alpha _3}$ are satisfied,
	${\hat \alpha _2}$ is the optimal value of $\alpha$.
	This is because, in this situation, based on a) - d), it is known that
	${\hat \alpha _2} < {\hat \alpha _1}$,
	${\hat \alpha _2}{\rho _1} = \left( {1 - {{\hat \alpha }_2}} \right){\rho _2}$,
	$ - {\varphi ^{\left( l \right)}}\frac{{d{E_{sum}}\left( {{{\hat \alpha }_2}} \right)}}{{d\alpha }} + {\rho _1} > 0$
	and
	$ - {\varphi ^{\left( l \right)}}\frac{{d{E_{sum}}\left( {\hat \alpha _2} \right)}}{{d\alpha }} - {\rho _2} \le 0$
	are satisfied.
	Based on the above and (\ref{stationarity2})-(\ref{primaryfeasible2}),
	when ${\alpha ^*} = {\hat \alpha _2}$,
	$1 \ge \lambda _1^* \ge 0$, $1 \ge \lambda _2^* \ge 0$, $\lambda _3^* = 0$ and $\lambda _4^* = 0$ are satisfied.
	Finally, the $\lambda _1^*$, $\lambda _2^*$, $\lambda _3^*$ and $\lambda _4^*$ corresponding to ${\alpha ^*} = {\hat \alpha _2}$ can be found to satisfy (\ref{stationarity}).
	
	\item[iii)] When ${\hat \alpha _1}{\rho _1} > \left( {1 - {{\hat \alpha }_1}} \right){\rho _2}$
	and
	${\hat \alpha _2} \in \left( {{\beta _1},{{\hat \alpha }_1}} \right)$,
	if ${\hat \alpha _3} > {\hat \alpha _2}$ is satisfied,
	${\hat \alpha _3}$ is the optimal value of $\alpha$.
	The reason why ${\alpha ^*} = {\hat \alpha _3}$ is that in this scenario, according to c) - e),
	${\hat \alpha _3}{\rho _1} > \left( {1 - {{\hat \alpha }_3}} \right){\rho _2}$,
	$ - {\varphi ^{\left( l \right)}}\frac{{d{E_{sum}}\left( {\hat \alpha _3} \right)}}{{d\alpha }} - {\rho _2} = 0$
	and
	${\hat \alpha _1} > {\hat \alpha _3}$
	are satisfied.
	Based on the above conditions and (\ref{stationarity2})-(\ref{primaryfeasible2}),
	when ${\alpha ^*} = {\hat \alpha _3}$, it is known that
	$\lambda _1^* = 0$, $\lambda _2^* = 1$, $\lambda _3^* = 0$ and $\lambda _4^* = 0$,
	and (\ref{stationarity}) is satisfied.
	
	\item[iv)] When ${\hat \alpha _1}{\rho _1} > \left( {1 - {{\hat \alpha }_1}} \right){\rho _2}$ and ${\hat \alpha _2} \notin \left( {{\beta _1},{{\hat \alpha }_1}} \right)$,
	if ${\beta _1} \ge {\hat \alpha _3}$ is satisfied, $\beta _1$ is the optimal value of $\alpha$.
	This is because in this case, according to c) and d),
	${\beta _1} \ge {\hat \alpha _2}$,
	${\beta _1}{\rho _1} \ge \left( {1 - {\beta _1}} \right){\rho _2}$
	and
	$ - {\varphi ^{\left( l \right)}}\frac{{d{E_{sum}}\left( {{\beta _1}} \right)}}{{d\alpha }} - {\rho _2} \le 0$
	are satisfied.
	Based on the above conditions and (\ref{stationarity2})-(\ref{primaryfeasible2}),
	when ${\alpha ^*} = {\beta _1}$, it is known that
	$\lambda _1^* = 0$, $\lambda _2^* = 1$, $\lambda _3^* \ge 0$ and $\lambda _4^* = 0$,
	and (\ref{stationarity}) is satisfied.
	
	\item[v)] When ${\hat \alpha _1}{\rho _1} > \left( {1 - {{\hat \alpha }_1}} \right){\rho _2}$ and ${\hat \alpha _2} \notin \left( {{\beta _1},{{\hat \alpha }_1}} \right)$,
	if ${\beta _1} < {\hat \alpha _3}$, which indicates ${\hat \alpha _3} > {\hat \alpha _2}$,
	similar to iii), ${\hat \alpha _3}$ is the optimal value of $\alpha$.
	
\end{enumerate}

\noindent
When ${\hat \alpha _1} \notin \left( {{\beta _1},1 - {\beta _2}} \right)$, the optimal value ${\alpha ^*}$ can be searched in a similar way to the case of ${\hat \alpha _1} \in \left( {{\beta _1},1 - {\beta _2}} \right)$.

According to the above, the proposed PDSA is summarized in \textbf{Algorithm \ref{A1}}.

\setcounter{equation}{33}
\begin{figure*}[t]
	\begin{equation}
		\begin{aligned}
			E_{sum}^{\sec } &= {\delta _{t,1}}\sum\limits_{n = 1}^{{N_1}} {\left( {{P_0}\left( {1 + \frac{{3{{\left\| {{{\bf{v}}_R}\left( n \right)} \right\|}^2}}}{{U_{tip}^2}}} \right) + \frac{{{\delta _{t,1}}{P_1}{v_0}}}{{\lambda \left( n \right)}} + \frac{1}{2}{d_0}\rho sA{{\left\| {{{\bf{v}}_R}\left( n \right)} \right\|}^3}} \right)}\\
			&+ {\delta _{t,2}}\sum\limits_{n = {N_1} + 1}^{N - 1} {\left( {{P_0}\left( {1 + \frac{{3{{\left\| {{{\bf{v}}_R}\left( n \right)} \right\|}^2}}}{{U_{tip}^2}}} \right) + \frac{{{\delta _{t,2}}{P_1}{v_0}}}{{\lambda \left( n \right)}} + \frac{1}{2}{d_0}\rho sA{{\left\| {{{\bf{v}}_R}\left( n \right)} \right\|}^3}} \right)}
			\label{Esumsec}
		\end{aligned}
	\end{equation}
	\hrulefill
\end{figure*}
\setcounter{equation}{28}

\emph{2) Transmit Power:} 
For the sub-problem given ${{\bf{Q}}_R}$ and $\alpha$,
the objective function is non-convex with respect to ${{\bf{P}}_S}$ and ${{\bf{P}}_{R,2}}$.
By introducing the slack variable ${\omega _2}$, the following equivalent problem is formed as
\begin{subequations}
	\begin{align}
		\mathcal{P}_{2.2.1}:\;&\mathop {\max }\limits_{{{\bf{P}}_S},{{\bf{P}}_{R,2}},{\omega _2}} {\omega _2} - {\varphi ^{\left( l \right)}}{E_{sum}} \label{P3.2.1a}\\
		{\mathrm{s.t.}}\;
		&{\phi _1}\sum\limits_{n = 1}^{{N_1}} {R_1^{\sec }\left( n \right)}  \ge {\omega _2}, \label{P3.2.1b}\\
		&{\phi _2}\sum\limits_{n = {N_1} + 1}^N {R_2^{\sec }\left( n \right)}  \ge {\omega _2}, \label{P3.2.1c}\\
		&(\textrm{\ref{P2b}}), (\textrm{\ref{P1c}}), (\textrm{\ref{P1d}}). \label{P3.2.1d}
	\end{align}
\end{subequations}

\noindent
Despite the non-convexity still present in (\ref{P3.2.1c}), an approximate convex form can be formed using the SCA method, namely (\ref{TPSCA}), shown at the top of this page, where $\left\{ {P_R^{\left( l \right)}\left( n \right),n \in {{\mathcal{N}}_2}} \right\}$ is fixed and obtained by \textbf{Algorithm \ref{A2}} in the $l$-th iteration. By replacing the corresponding part of constraint (\ref{P3.2.1c}) with the approximate form, an approximate convex problem of $\mathcal{P}_{2.2.1}$ that can be solved using the CVX toolbox is modeled as \cite{BoydS2004Book}
\setcounter{equation}{30}
\begin{subequations}
	\begin{align}
		\mathcal{P}_{2.2.2}:\;&\mathop {\max }\limits_{{{\bf{P}}_S},{{\bf{P}}_{R,2}},{\omega _2}} {\omega _2} - {\varphi ^{\left( l \right)}}{E_{sum}} \label{P3.2.2a}\\
		{\mathrm{s.t.}}\;
		&{\phi _2}\sum\limits_{n = {N_1} + 1}^N {\left( {R_D^{\sec }\left( n \right) - K\left( n \right)} \right)}  \ge {\omega _2}, \label{P3.2.2b}\\
		&(\textrm{\ref{P3.2.1b}}), (\textrm{\ref{P3.2.1d}}), \label{P3.2.2c}
	\end{align}
\end{subequations}
where $R_D^{\sec }\left( n \right) = B{\log _2}\left( {{\sigma ^2} + {P_R}\left( n \right){{\left| {{h_{RD}}\left( n \right)} \right|}^2}} \right)$.

\emph{3) Trajectory of $R$:} By reducing problem $\mathcal{P}_{2}$, the sub-problem with respect to ${{{\bf{Q}}_R}}$ is expressed as
\begin{subequations}
	\begin{align}
		\mathcal{P}_{2.3.1}:\;&\mathop {\max }\limits_{{{\bf{Q}}_R}} R_{ave}^{\sec } - {\varphi ^{\left( l \right)}}{E_{sum}} \label{P3.3.1a}\\
		{\mathrm{s.t.}}\;
		&(\textrm{\ref{P2b}}), (\textrm{\ref{P1e}}) - (\textrm{\ref{P1g}}). \label{P3.3.1b}
	\end{align}
\end{subequations}
Even with given ${{\bf{P}}_S}$, ${{\bf{P}}_{R,2}}$ and ${\alpha }$, problem $\mathcal{P}_{2.3.1}$ is still intractable to solve optimally since non-convexity and strong coupling with respect to ${{\bf{Q}}_R}$ still exist in objective function and constraint (\ref{P2b}).
To tackle the strong coupling with respect to ${{\bf{Q}}_R}$, slack variables ${\omega _3}$,
$\lambda  = \left\{ {\lambda \left( n \right),n \in {\mathcal{N}}} \right\}$,
$\nu  = \left\{ {\nu \left( n \right),n \in {{\mathcal{N}}_1}} \right\}$,
$\kappa  = \left\{ {\kappa \left( n \right),n \in {{\mathcal{N}}_2}} \right\}$,
$\varpi  = \left\{ {\varpi \left( n \right),n \in {{\mathcal{N}}_2}} \right\}$,
and
$\vartheta  = \left\{ {\vartheta \left( n \right),n \in {{\mathcal{N}}_1}} \right\}$
are introduced to obtain a equivalent form of problem $\mathcal{P}_{2.3.1}$, which is expressed as
\begin{subequations}\small
	\begin{align}
		\mathcal{P}_{2.3.2}:\;&\mathop {\max }\limits_{{{\bf{Q}}_R},{\omega _3},\lambda ,\nu ,\kappa ,\varpi ,\vartheta } {\omega _3} - {\varphi ^{\left( l \right)}}E_{sum}^{\sec } \label{P3.3.2a}\\
		{\mathrm{s.t.}}\;
		&{\lambda ^2}\left( n \right) \le {\left\| {{{\bf{q}}_R}\left( n \right) - {{\bf{q}}_R}\left( {n - 1} \right)} \right\|^2},n \in {\mathcal{N}}, \label{P3.3.2b}\\
		&{{\phi}_1}\sum\limits_{n = 1}^{{N_1}} {R_1^{\sec ,1}\left( n \right)}  \ge {\omega _3}, \label{P3.3.2c}\\
		&{\phi _2}\sum\limits_{n = {N_1} + 1}^N {R_2^{\sec ,1}\left( n \right)}  \ge {\omega _3}, \label{P3.3.2d}\\
		&{P_S}\left( n \right){\left| {\bar h_{SW}^{\sec }} \right|^2} \le \frac{{{\varepsilon}{\beta _0} {\hat P_R^J}}}{{\vartheta \left( n \right)}}, n \in {{\mathcal{N}}_1}, \label{P3.3.2e}\\
		&\nu \left( n \right) \ge {\left\| {{{\bf{q}}_R}\left( n \right) - {{\bf{q}}_S}} \right\|^2} + {H^2},n \in {{\mathcal{N}}_1}, \label{P3.3.2f}\\
		&\kappa \left( n \right) \ge {\left\| {{{\bf{q}}_R}\left( n \right) - {{\bf{q}}_D}} \right\|^2} + {H^2},n \in {{\mathcal{N}}_2}, \label{P3.3.2g}\\
		&\varpi \left( n \right) \le {\left( {\left\| {{{\bf{q}}_R}\left( n \right) - {{{\bf{\hat q}}}_E}} \right\| - {r_E}} \right)^2} + {H^2},n \in {{\mathcal{N}}_2}, \label{P3.3.2h}\\
		&\vartheta \left( n \right) \ge {\left( {\left\| {{{\bf{q}}_R}\left( n \right) - {{{\bf{\hat q}}}_W}} \right\| + {r_W}} \right)^2} + {H^2},n \in {{\mathcal{N}}_1}, \label{P3.3.2i}\\
		&(\textrm{\ref{P1e}}) - (\textrm{\ref{P1g}}), \label{P3.3.2j}
	\end{align}
\end{subequations}
where $E_{sum}^{\sec }$ is given in (\ref{Esumsec}), shown at the top of this page,
$R_1^{\sec ,1}\left( n \right) = B{\log _2}\left( {1 + \frac{{{P_S}\left( n \right){\beta _0}}}{{\nu \left( n \right)\left( {P_R^J\psi  + {\sigma ^2}} \right)}}} \right)$,
$R_2^{\sec ,1}\left( n \right) = R_D^{\sec ,1}\left( n \right) - R_E^{\sec ,1}\left( n \right)$,
$R_D^{\sec ,1}\left( n \right) = B{\log _2}\left( {1 + \frac{{{P_R}\left( n \right){\beta _0}}}{{\kappa \left( n \right){\sigma ^2}}}} \right)$,
and
$R_E^{\sec ,1}\left( n \right) = B{\log _2}\left( {1 + \frac{{{P_R}\left( n \right){\beta _0}}}{{\varpi \left( n \right){\sigma ^2}}}} \right)$.
Subsequently, to handle the residual non-convexity, the first-order Taylor expansion method is applied to build an approximate convex problem which can be efficiently solved using the CVX toolbox \cite{BoydS2004Book}, namely
\setcounter{equation}{34}
\begin{subequations}
	\begin{align}
		\mathcal{P}_{2.3.3}:\;&\mathop {\max }\limits_{{{\bf{Q}}_R},{\omega _3},\lambda ,\nu ,\kappa ,\varpi ,\vartheta } {\omega _3} - {\varphi ^{\left( l \right)}}E_{sum}^{\sec } \label{P3.3.3a}\\
		{\mathrm{s.t.}}\;
		&{\lambda ^2}\left( n \right) \le B\left( n \right),n \in {\mathcal{N}}, \label{P3.3.3b}\\
		&{{\phi}_1}\sum\limits_{n = 1}^{{N_1}} {\left( {R_1^{\sec ,2}\left( n \right) - C\left( n \right)} \right)}  \ge {\omega _3}, \label{P3.3.3c}\\
		&{{\phi}_2}\sum\limits_{n = {N_1} + 1}^N {\left( {R_2^{\sec ,2}\left( n \right) - D\left( n \right) - E\left( n \right)} \right)} \ge {\omega _3}, \label{P3.3.3d}\\
		&\varpi \left( n \right) \le F\left( n \right) + {H^2},n \in {{\mathcal{N}}_2}, \label{P3.3.3e}\\
		&(\textrm{\ref{P3.3.2e}}), (\textrm{\ref{P3.3.2f}}), (\textrm{\ref{P3.3.2g}}), (\textrm{\ref{P3.3.2i}}), (\textrm{\ref{P3.3.2j}}), \label{P3.3.3f}
	\end{align}
\end{subequations}
where
$R_1^{\sec ,2}\left( n \right) = B{\log _2}\left( {\nu \left( n \right)\left( {P_R^J\psi  + {\sigma ^2}} \right) + {P_S}\left( n \right){\beta _0}} \right)$,
$R_2^{\sec ,2}\left( n \right) = B{\log _2}\left( {\kappa \left( n \right){\sigma ^2} + {P_R}\left( n \right){\beta _0}} \right) + {\log _2}\left( {\varpi \left( n \right){\sigma ^2}} \right)$,
and $B\left( n \right), \cdots ,F\left( n \right)$ are presented as (\ref{SCApart}), shown at the top of the next page, where
$\left\{ {{\bf{q}}_R^{\left( l \right)}\left( n \right),n \in {\mathcal{N}}} \right\}$,
$\left\{ {{\nu ^{\left( l \right)}}\left( n \right),n \in {{\mathcal{N}}_1}} \right\}$,
$\left\{ {{\kappa ^{\left( l \right)}}\left( n \right),n \in {{\mathcal{N}}_2}} \right\}$,
and
$\left\{ {{\varpi ^{\left( l \right)}}\left( n \right),n \in {{\mathcal{N}}_2}} \right\}$
are obtained by \textbf{Algorithm \ref{A2}} in $l$-th iteration.

Finally, the AO-based algorithm, summarized in \textbf{Algorithm \ref{A2}}, is developed to solve problem $\mathcal{P}_{1}$ sub-optimally, where $\hat \xi$ represents the accuracy of convergence.

\begin{algorithm}[!tb]
	\caption{Proposed AO-Based Algorithm for Non-Convex Problem $\mathcal{P}_{1}$}
	\label{A2}
	\KwIn{Initialize ${\bf{P}}_S^{\left( 0 \right)}$, ${\bf{P}}_{R,2}^{\left( 0 \right)}$, ${\bf{Q}}_R^{\left( 0 \right)}$ and ${\alpha ^{\left( 0 \right)}}$.
		Calculate ${\varphi ^{\left( 0 \right)}}$ and $l \leftarrow 0$;}
	\Do{${\varphi ^{\left( l \right)}} - {\varphi ^{\left( {l - 1} \right)}} > \hat \xi $}
	{
		Solve $\mathcal{P}_{2.1.2}$ by \textbf{Algorithm 1} with given ${\bf{P}}_S^{\left( l \right)}$, ${\bf{P}}_{R,2}^{\left( l \right)}$ and ${\bf{Q}}_R^{\left( l \right)}$;\\
		Solve $\mathcal{P}_{2.2.2}$ with given ${\bf{Q}}_R^{\left( l \right)}$ and ${\alpha ^{\left( l + 1 \right)}}$;\\
		Solve $\mathcal{P}_{2.3.3}$ with given ${\bf{P}}_S^{\left( {l + 1} \right)}$, ${\bf{P}}_{R,2}^{\left( {l + 1} \right)}$ and ${\alpha ^{\left( {l + 1} \right)}}$;\\
		$l \leftarrow l + 1$;\\
		Calculate ${\varphi ^{\left( l \right)}}$;
	}
	\KwOut{${\varphi ^{\left( l \right)}}$ with ${\bf{P}}_S^* = {\bf{P}}_S^{\left( l \right)}$, ${\bf{P}}_{R,2}^* = {\bf{P}}_{R,2}^{\left( l \right)}$, ${\bf{Q}}_R^* = {\bf{Q}}_R^{\left( l \right)}$, ${\alpha ^*} = {\alpha ^{\left( l \right)}}$.}
\end{algorithm}

\subsection{Convergence and Complexity Analysis}

In this subsection, based on the results presented in the previous subsection, the convergence of \textbf{Algorithm \ref{A2}} is proved as follows.
\begin{proof}
	Suppose $\varphi \left( {{\bf{P}}_S^{\left( {l - 1} \right)},{\bf{P}}_{R,2}^{\left( {l - 1} \right)},{\bf{Q}}_R^{\left( {l - 1} \right)},{\alpha ^{\left( {l - 1} \right)}}} \right)$ indicates the objective value of the problem $\mathcal{P}_{1}$ in $\left( l - 1 \right)$-th iteration.
	First, since the optimal value of problem $\mathcal{P}_{2.1.1}$ is obtained by solving the equivalent form $\mathcal{P}_{2.1.2}$ via PDSA, we have
	\setcounter{equation}{36}
	\begin{equation}
		\begin{aligned}
			&\varphi \left( {{\bf{P}}_S^{\left( {l - 1} \right)},{\bf{P}}_{R,2}^{\left( {l - 1} \right)},{\bf{Q}}_R^{\left( {l - 1} \right)},{\alpha ^{\left( {l - 1} \right)}}} \right) \\
			&\;\;\;\;\;\;\;\;\;\;\;\;\;\;\;\;\le \varphi \left( {{\bf{P}}_S^{\left( {l - 1} \right)},{\bf{P}}_{R,2}^{\left( {l - 1} \right)},{\bf{Q}}_R^{\left( {l - 1} \right)},{\alpha ^{\left( l \right)}}} \right).
			\label{}
		\end{aligned}
	\end{equation}
	Second, by solving the approximate form $\mathcal{P}_{2.2.2}$, a sub-optimal value for problem $\mathcal{P}_{2.2.1}$ is achieved, yielding
	\begin{equation}
		\begin{aligned}
			&\varphi \left( {{\bf{P}}_S^{\left( {l - 1} \right)},{\bf{P}}_{R,2}^{\left( {l - 1} \right)},{\bf{Q}}_R^{\left( {l - 1} \right)},{\alpha ^{\left( l \right)}}} \right) \\
			&\;\;\;\;\;\;\;\;\;\;\;\;\;\;\;\;\le \varphi \left( {{\bf{P}}_S^{\left( l \right)},{\bf{P}}_{R,2}^{\left( l \right)},{\bf{Q}}_R^{\left( {l - 1} \right)},{\alpha ^{\left( l \right)}}} \right).
			\label{}
		\end{aligned}
	\end{equation}
	Similarly, $\varphi \left( {{\bf{P}}_S^{\left( l \right)},{\bf{P}}_{R,2}^{\left( l \right)},{\bf{Q}}_R^{\left( {l - 1} \right)},{\alpha ^{\left( l \right)}}} \right) \le \varphi \left( {{\bf{P}}_S^{\left( l \right)},{\bf{P}}_{R,2}^{\left( l \right)},{\bf{Q}}_R^{\left( l \right)},{\alpha ^{\left( l \right)}}} \right)$ is established by resolving problem $\mathcal{P}_{2.3.3}$.
	
	Based on the above, it is obvious that the objective function $\varphi$ is always non-decreasing after each iteration.
	Additionally, $\varphi$ is upper bounded by a finite value.
	Conclusively, \textbf{Algorithm \ref{A2}} is convergent.
\end{proof}
Furthermore, the overall computational complexity of \textbf{Algorithm \ref{A2}} is analyzed as follows. In each iteration of \textbf{Algorithm \ref{A2}}, the problem $\mathcal{P}_{2.1.2}$ is optimized via PDSA, whose complexity is the same as a bisection search method, and the problems $\mathcal{P}_{2.2.2}$ and $\mathcal{P}_{2.3.3}$ are sequentially solved by existing standard convex solvers. Consequently, the individual complexities are given by $\mathcal{O}\left( {{{\log }_2}\left( {\frac{1}{\varsigma }} \right)} \right)$, $\mathcal{O}\left( {{N^{3.5}}\log \left( {\frac{1}{\varsigma }} \right)} \right)$, and $\mathcal{O}\left( {{{\left( {5N} \right)}^{3.5}}\log \left( {\frac{1}{\varsigma }} \right)} \right)$, respectively, where $\varsigma$ denotes the accuracy \cite{WangW2021JSAC}. Thus, the total computational complexity of \textbf{Algorithm \ref{A2}} is ${\cal O}\left( {{N_{num}}\left( {{\left( {5N} \right)}^{3.5}} + {{N}^{3.5}} \right)\log \left( {\frac{1}{\varsigma }} \right)} \right)$, where ${N_{num}}$ is the required iteration number.

\setcounter{equation}{35}
\begin{figure*}[t]
	\begin{subequations}
		\begin{align}
			&B\left( n \right) = {\left\| {{\bf{q}}_R^{\left( l \right)}\left( n \right) - {\bf{q}}_R^{\left( l \right)}\left( {n - 1} \right)} \right\|^2} + 2{\left( {{\bf{q}}_R^{\left( l \right)}\left( n \right) - {\bf{q}}_R^{\left( l \right)}\left( {n - 1} \right)} \right)^T} \left( {{{\bf{q}}_R}\left( n \right) - {{\bf{q}}_R}\left( {n - 1} \right) - {\bf{q}}_R^{\left( l \right)}\left( n \right) + {\bf{q}}_R^{\left( l \right)}\left( {n - 1} \right)} \right) \label{}\\
			&C\left( n \right) = B{\log _2}\left( {{\nu ^{\left( l \right)}}\left( n \right)\left( {P_R^J\psi  + {\sigma ^2}} \right)} \right) + \frac{B}{{{\nu ^{\left( l \right)}}\left( n \right)\ln \left( 2 \right)}}\left( {\nu \left( n \right) - {\nu ^{\left( l \right)}}\left( n \right)} \right) \label{}\\
			&D\left( n \right) = B{\log _2}\left( {{\kappa ^{\left( l \right)}}\left( n \right){\sigma ^2}} \right) + \frac{B}{{{\kappa ^{\left( l \right)}}\left( n \right)\ln \left( 2 \right)}}\left( {\kappa \left( n \right) - {\kappa ^{\left( l \right)}}\left( n \right)} \right) \label{}\\
			&E\left( n \right) = B{\log _2}\left( {{\varpi ^{\left( l \right)}}\left( n \right){\sigma ^2} + {P_R}\left( n \right){\beta _0}} \right) + \frac{{\left( {\varpi \left( n \right) - {\varpi ^{\left( l \right)}}\left( n \right)} \right)B{\sigma ^2}}}{{\left( {{\varpi ^{\left( l \right)}}\left( n \right){\sigma ^2} + P_R^{\left( l \right)}\left( n \right){\beta _0}} \right)\ln \left( 2 \right)}} \label{}\\
			&F\left( n \right) = {\left( {\left\| {{\bf{q}}_R^{\left( l \right)}\left( n \right) - {{{\bf{\hat q}}}_E}} \right\| - {r_E}} \right)^2} + \frac{2{\left( {\left\| {{\bf{q}}_R^{\left( l \right)}\left( n \right) - {{{\bf{\hat q}}}_E}} \right\| - {r_E}} \right)}}{{\left\| {{\bf{q}}_R^{\left( l \right)}\left( n \right) - {{{\bf{\hat q}}}_E}} \right\|}} {\left( {{\bf{q}}_R^{\left( l \right)}\left( n \right) - {{{\bf{\hat q}}}_E}} \right)^T}\left( {{{\bf{q}}_R}\left( n \right) - {\bf{q}}_R^{\left( l \right)}\left( n \right)} \right) \label{}
		\end{align} \label{SCApart}
	\end{subequations}
	\hrulefill
\end{figure*}

\section{Simulation Results and Discussion}
\label{sec:Simulation}

\begin{table}[t]
		\caption{\textit{List of Simulation Parameters}}
		\label{Table 2}
		\begin{center}
			\begin{tabular}{c|c | c |c }
				\Xhline{1.2pt}
				\textbf{Notation}   	& \textbf{Value} & \textbf{Notation}   	& \textbf{Value}\\
				\hline
				${{\mathbf{q}}^I}$, ${{\mathbf{q}}^F}$      & $\left[ {_{350,}^{0,}}{_{350}^{700}} \right]$ &	${{\mathbf{q}}_{S}}$, ${{\mathbf{q}}_{D}}$    & $\left[ {_{500,}^{200,}}{_{200}^{500}} \right]$ \\
				\hline
				${{\mathbf{\hat{q}}}_{W}}$, ${{\mathbf{\hat{q}}}_{E}}$    & $\left[ {_{450,}^{350,}}{_{200}^{350}} \right]$&
				$r_j$                 	& $15$ m\\
				\hline
				$H$  					& $75$ m &
				${\sigma ^2}$			& ${{-120}}$ dB\\
				\hline
				${P_{S,\max}}$			& $1$ W &
				${P_{R,\max}}$			& $1$ W\\
				\hline
				${\hat P_{R}^{J}}$			& $1$ W &
				${V_{\max}}$			& $50$ m/s\\
				\hline
				${N_1}$					& $50$ &
				${N_2}$					& $50$\\
				\hline
				${\beta_0}$				& ${-30}$ dB &
				$\zeta$					& $2.1$\\
				\hline
				$\varepsilon$			& $0.01$&
				$\hat{\xi}$				& $0.01$\\
				\Xhline{1.2pt}
			\end{tabular}
		\end{center}
\end{table}

This section presents simulation results that demonstrate the efficacy of the proposed algorithm, which is referred to as 'Prop' in the following description. It is hereby stated that the parameters for the propulsion power of $R$ are based on \cite{ZengY2019TWC}, and the particulars of the remaining parameter setup are listed in Table \ref{Table 2} unless otherwise specified. To demonstrate the superiority of the proposed scheme, the following benchmark schemes are considered.

\begin{enumerate}
	
	\item \textit{Benchmark 1:} The trajectory of $R$ and the transmit power are optimized with a given phase-switching factor, referred to as 'Ben1' in the following description.
	
	\item \textit{Benchmark 2:} $R$ works with a fixed trajectory, while the transmit power and the phase-switching factor are optimized, referred to as 'Ben2' in the following description.
	
\end{enumerate}

\begin{figure}[t]
	\centering
	\includegraphics[width = 0.3  \textwidth]{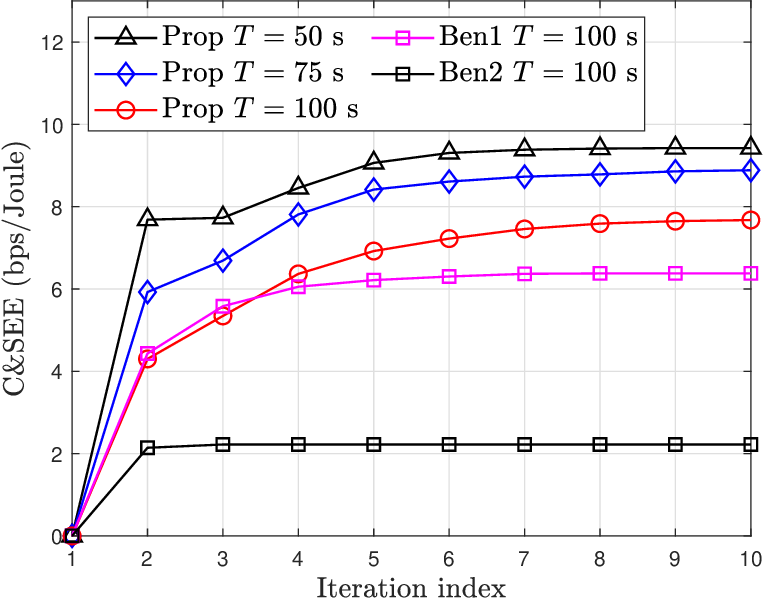}
	\caption{C\&SEE versus the number of iterations for different flying periods $T$ and schemes.}
	\label{fig02}
\end{figure}
The behavior of \textbf{Algorithm \ref{A2}} for various flying periods is shown in Fig. \ref{fig02}.
The convergence is deemed to be attained when the increase of C\&SEE is less than $\hat{\xi} = 0.01$.
As shown in Fig. \ref{fig02}, C\&SEE increases rapidly before plateauing as the number of iterations grows, reaching convergence within 10 iterations.
Additionally, observation shows that Prop's performance significantly exceeds that of other benchmark schemes, demonstrating its superiority.

\begin{figure}[t]
	\centering
	\subfigure[The trajectory of $R$.]{
		\label{fig03a}
		\includegraphics[width = 0.3  \textwidth]{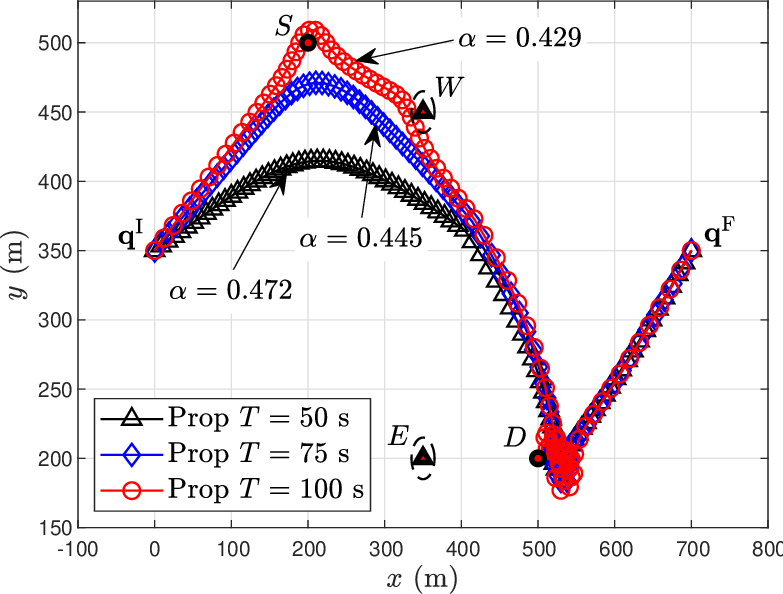}}
	\subfigure[The transmit power of $S$.]{
		\label{fig03b}
		\includegraphics[width = 0.3  \textwidth]{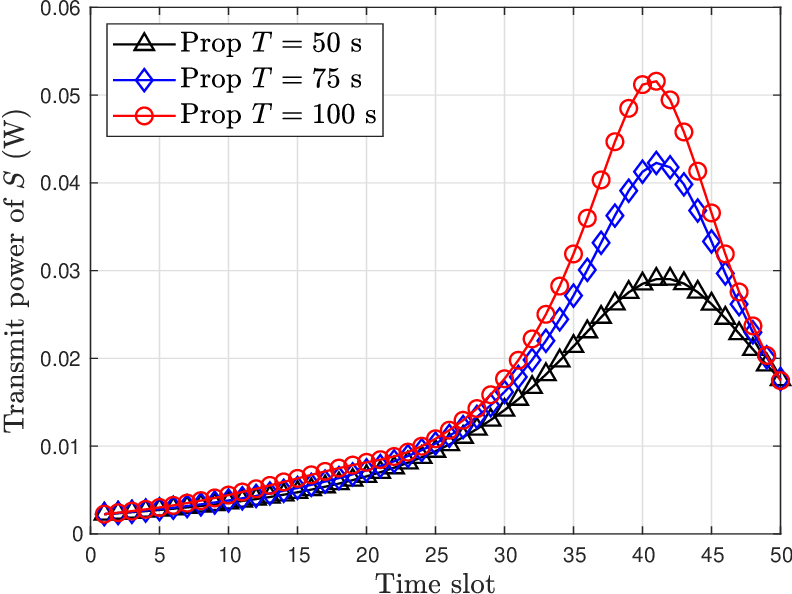}}
	\caption{Optimized trajectory and transmit power with different flying periods $T$.}
	\label{fig03}
\end{figure}
The optimal trajectory and transmit power with different flying periods $T$ are depicted in Fig. \ref{fig03}, where $T$ is equal to $50$ s, $70$ s, $100$ s, and $120$ s. Fig. \ref{fig03a} clearly shows that in the first phase, as $T$ increases, $R$ gains greater maneuverability to approach the ground units ($S$ and $W$), leading to enhanced air-ground channels for both receiving and jamming, which in turn enables $S$ to transmit more power. In the second phase, it is typical for $R$ to maneuver away from $E$ to avoid wiretapping. Moreover, an increase in $T$ leads to a decrease in optimal phase-switching factor $\alpha$. This is due to the introduction of collaborative interference, where the first phase of the system achieves higher performance compared to the second phase. To maintain optimal AC\&SR under the DF protocol, $\alpha$ is set to a smaller value as $T$ increases. This indicates that $\alpha$ is a crucial factor to consider in decision-making.

\begin{figure}[t]
	\centering
	\subfigure[The trajectory of $R$.]{
		\label{fig04a}
		\includegraphics[width = 0.3  \textwidth]{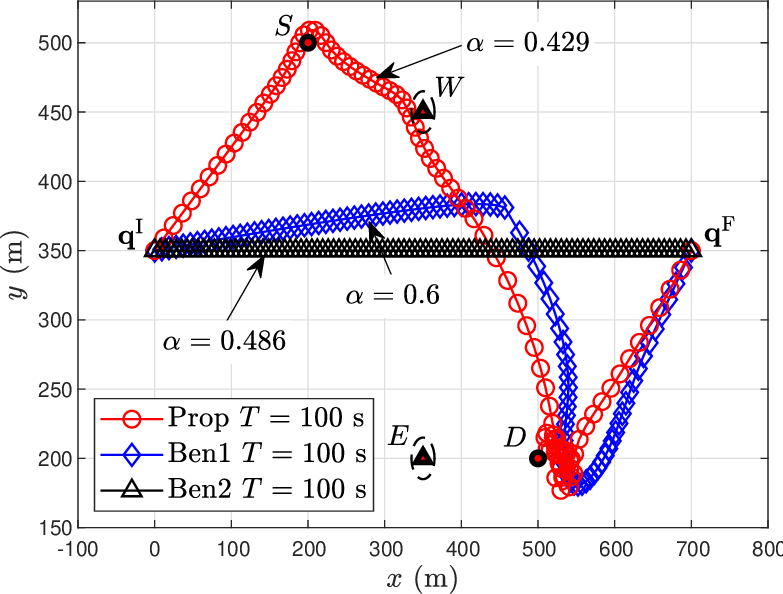}}
	\subfigure[The transmit power of $S$.]{
		\label{fig04b}
		\includegraphics[width = 0.3  \textwidth]{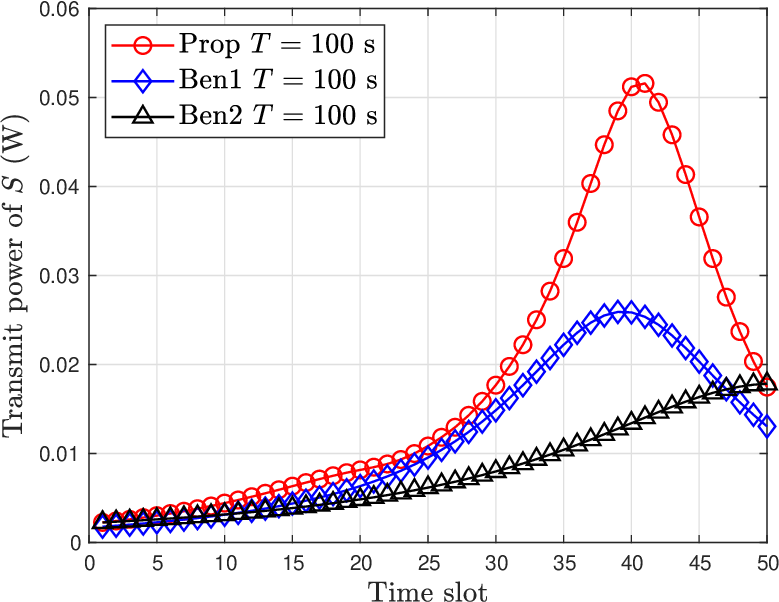}}
	\caption{Trajectory and transmit power with different schemes.}
	\label{fig04}
\end{figure}
Fig \ref{fig04} further illustrates the trajectories and transmission power under different strategies. It is evident that in Ben1, since the phase-switching factor $\alpha$ is no longer optimized based on the environment, $R$ is forced to approach $D$ prematurely before the end of the first phase to ensure adequate transmission quality in the second phase. This approach leads to insufficient interference to $W$ in the first phase and creates a time constraint in the second phase. Furthermore, although Ben2 involves an optimized decision for $\alpha$, excluding trajectory optimization greatly diminishes its overall effectiveness. These findings underscore the importance of jointly optimizing $\alpha$ and $R$'s trajectory to maximize C\&SEE.

\begin{figure}[t]
	\centering
	\includegraphics[width = 0.3  \textwidth]{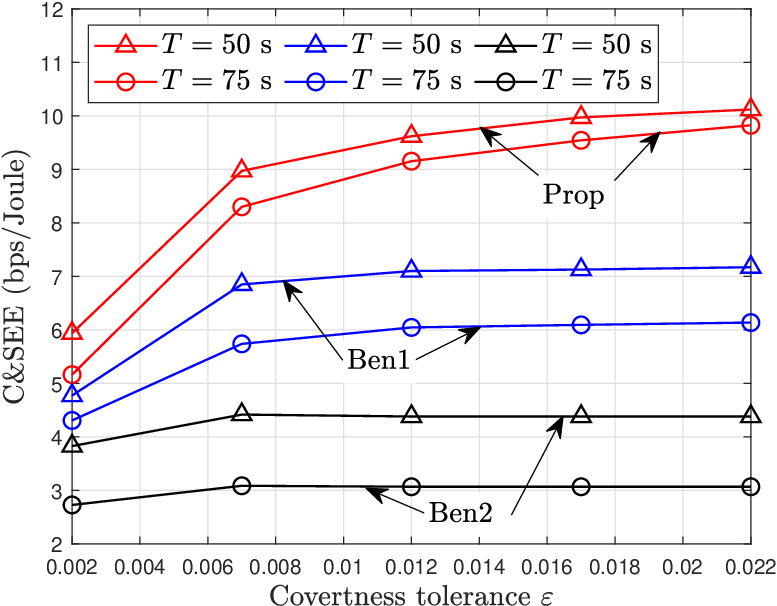}
	\caption{The impact of varying covertness tolerance $\varepsilon$.}
	\label{fig05}
\end{figure}
To demonstrate the effect of varying covertness tolerance $\varepsilon$, the curve of C\&SEE with different benchmarks is shown in Fig. \ref{fig05}. As the value of $\varepsilon$ increases, the graphs corresponding to Prop, Ben1, and Ben2 display an upward trend. This is because when the tolerance level goes up, it becomes easier to meet the requirement for covertness, which enables the transmitter $S$ to transmit higher power. Meanwhile, as $\varepsilon$ exceeds a certain degree, the trend of increasing C\&SEE starts to level off, particularly in the curve displayed in Ben2. This suggests that merely increasing covertness tolerance does not always result in substantial C\&SEE gains. Instead, it can expose confidential information to potential threats.

\begin{figure}[t]
	\centering
	\includegraphics[width = 0.3  \textwidth]{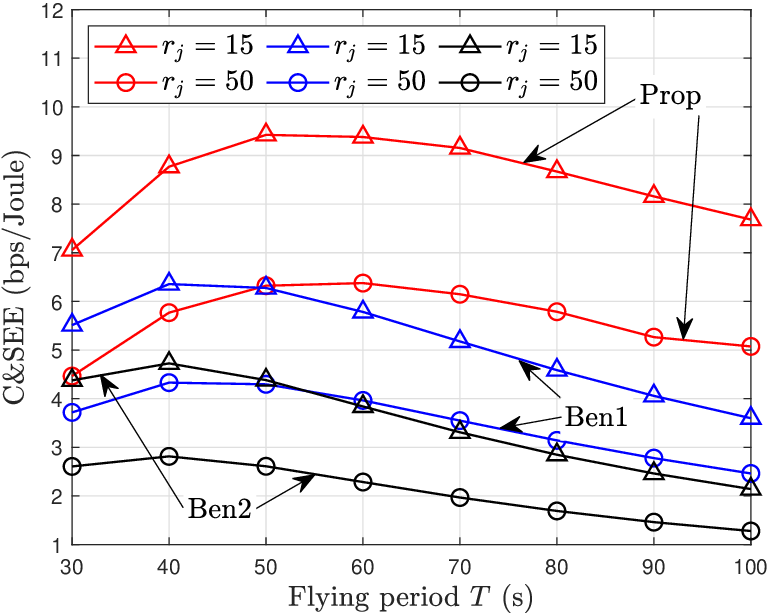}
	\caption{C\&SEE versus varying flying period $T$.}
	\label{fig06}
\end{figure}
For two different observation frequencies of $W$, Fig. \ref{fig06} illustrates the trend in C\&SEE with an increasing flight period $T$. Unlike the relationship shown in Fig. \ref{fig05} between C\&SEE and covertness tolerance, for an increased flight period, C\&SEE follows a different pattern. Initially, C\&SEE increases, then begins to decline after reaching a certain point, and eventually flattens out. The underlying reason is that the initial gain in AC\&SR resulting from increased flight time outweighs the associated energy loss significantly. However, beyond a certain threshold, the gain in AC\&SR no longer offsets the energy loss, leading to a decline in C\&SEE. This phenomenon highlights the criticality of the selection of flight period $T$ to achieve optimal C\&SEE performance.

\begin{figure}[t]
	\centering
	\includegraphics[width = 0.3  \textwidth]{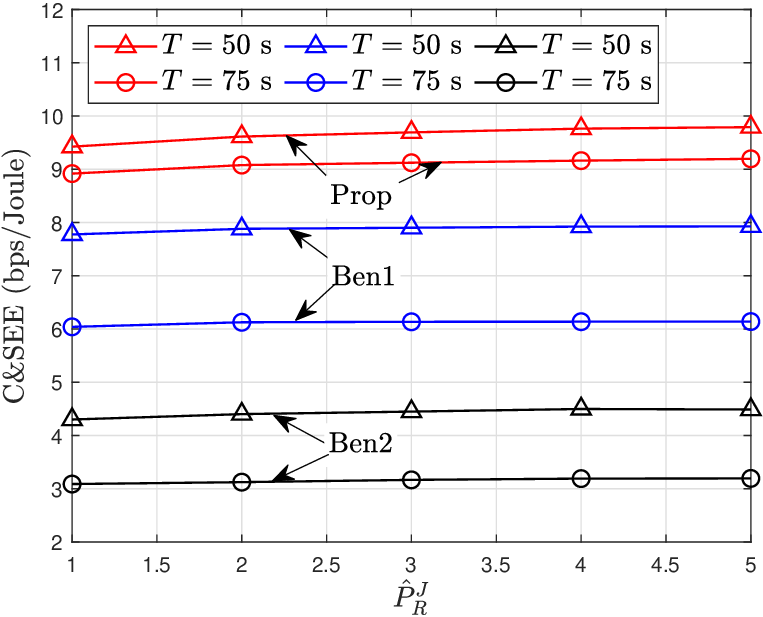}
	\caption{The effect of varying ${\hat P_R^J}$ on C\&SEE.}
	\label{fig07}
\end{figure}
Fig. \ref{fig07} examines how the uniform distribution interval, $\left[ {0,\hat P_R^J} \right]$, affects C\&SEE.
It is evident that when $\hat P_R^J$ rises, C\&SEE has a increasing trend.
This occurs because higher $\hat P_R^J$ values introduce stronger confounding and bias into $W$, leading to more pronounced misleading effects.
Under the aforementioned conditions while maintaining equivalent communication covertness, node $S$ can enhance its transmission power, consequently improving the C\&SEE performance.

\begin{figure}[t]
	\centering
	\includegraphics[width = 0.3  \textwidth]{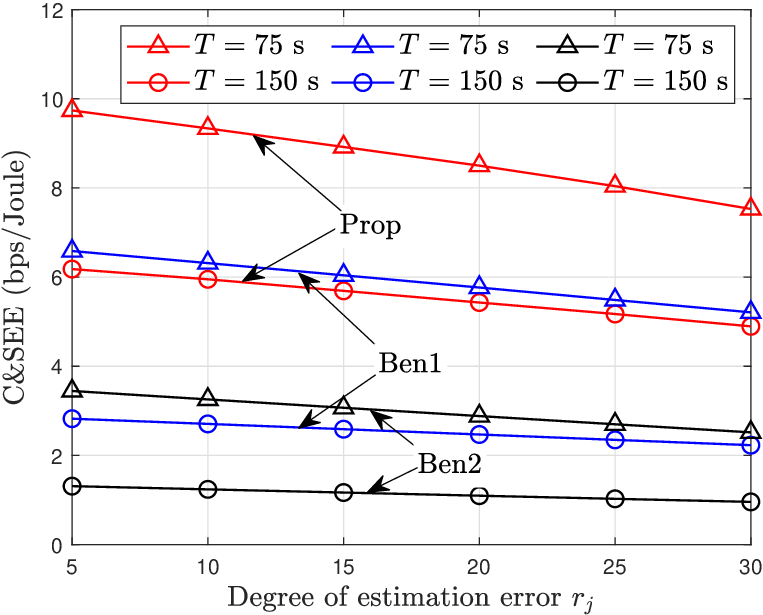}
	\caption{The impact of varying degree of estimation error $r_j$.}
	\label{fig08}
\end{figure}
The relationship between the degree of estimation error $r_j$ and C\&SEE is illustrated in Fig. \ref{fig08}.
It is evident that, regardless of the strategy employed, C\&SEE exhibits a monotonically decreasing trend with respect to $r_j$.
This phenomenon arises because, as the degree of estimation error increases, $R$ is compelled to expand its vigilance zone to prevent unexpected incidents.
Consequently, $R$ expends additional energy on extensive detours and rerouting, resulting in a downward trend in C\&SEE.
Ultimately, Prop outperforms Ben1 and Ben2 in terms of C\&SEE for all parameters considered in Figs. \ref{fig05} - \ref{fig08}, demonstrating its superiority.

\section{Conclusion and Future Work}
\label{sec:Conclusion}

In this work, a UAV relay assisted cooperative jamming covert and secure communication framework that considers the practical case of uncertain malicious nodes' location was investigated. By designing the trajectory of the UAV, the transmitter's power, and the phase-switching factor to maximize C\&SEE, a robust FP optimization problem was established. To deal with this multivariate coupled non-convex FP optimization problem, an efficient algorithm was proposed based on AO, primal-dual search, and SCA methods to obtain a sub-optimal solution. Numerical results exhibited the efficiency and superiority of the proposed algorithm and each parameter's impact on C\&SEE. Finally, as shown in Fig. \ref{fig06} of the numerical results, an optimal flight period exists within the proposed scheme. Exploring methods to determine this presents a valuable direction for future research.

\end{document}